\pdfoutput=1

\documentclass[sigconf]{acmart}

\usepackage{cleveref}
\usepackage{tikz}
\usetikzlibrary{positioning}
\usepackage{subcaption}
\usepackage{eurosym}
\usepackage{acmart-taps}

\AtBeginDocument{%
  }

\copyrightyear{2026}
\acmYear{2026}
\setcopyright{cc}
\setcctype{by}
\acmConference[BuildSys '26]{The 13th ACM International Conference on Systems for Energy-Efficient Buildings, Cities, and Transportation}{June 22--25, 2026}{Banff, AB, Canada}
\acmBooktitle{The 13th ACM International Conference on Systems for Energy-Efficient Buildings, Cities, and Transportation (BuildSys '26), June 22--25, 2026, Banff, AB, Canada}
\acmDOI{10.1145/3744256.3812575}
\acmISBN{979-8-4007-2012-3/2026/06}

\begin{document}

\title{Should Belgians Install an Advanced Energy Management System in Their Electrified Dwelling ?}

\author{Julian Ruddick}
\email{julian.ruddick@vub.be}
\orcid{0000-0002-9063-6939}
\affiliation{%
  \institution{Vrije Universiteit Brussel}
  \city{Brussels}
  \country{Belgium}
}

\author{Cedric De Cauwer}
\email{cedric.de.cauwer@vub.be}
\orcid{0000-0001-8214-390X}
\affiliation{%
  \institution{Vrije Universiteit Brussel}
  \city{Brussels}
  \country{Belgium}
}

\author{Thierry Coosemans}
\email{thierry.coosemans@vub.be}
\orcid{0000-0002-7732-595X}
\affiliation{%
  \institution{Vrije Universiteit Brussel}
  \city{Brussels}
  \country{Belgium}
}

\author{Maarten Messagie}
\email{maarten.messagie@vub.be}
\orcid{0000-0003-1733-9572}
\affiliation{%
  \institution{Vrije Universiteit Brussel}
  \city{Brussels}
  \country{Belgium}
}

\renewcommand{\shortauthors}{Ruddick et al.}

\begin{abstract}
As dwellings electrify and more complex electricity pricing spreads in Belgium, the question arises: What are the potential benefits of installing an advanced energy management systems in an electrified Belgian dwelling ?

To answer this question, we simulate 500 representative Belgian dwellings in which fossil systems are replaced by controllable electric technologies: space heating via air-to-air heat pumps, domestic hot water via electric boilers, and private transport via electric vehicles with home chargers. Photovoltaic systems are assigned according to current adoption rates.

We evaluate two advanced energy management systems (EMS): model predictive control (MPC) and TreeC, a machine learning method using decision trees to model the EMS strategy. These are compared to a rule-based controller (RBC) that mimics default operation and an MPC with perfect forecasts (MPC P) as an upper bound. Performance for these EMSs is assessed for 2023 under a real contract with hourly prices and a monthly peak tariff.

Average annual savings versus RBC are 82.5 €, 116.6 €, and 177.4 € for TreeC, MPC, and MPC P, respectively. All EMSs deliver similar thermal comfort (26–31 Kh). MPC saves more, while TreeC achieves higher EV charge completion (93.3\% vs 88.8\% for MPC, with a max charge of 95\%) and a smoother aggregate load across the 500 dwellings. Savings remain modest relative to EMS purchase (280–680 €) and subscription (0–360 €/year) costs which is not favourable towards widespread adoption.

The data and code to reproduce this study are available at \url{https://github.com/EVERGi/should_belgians_install_advanced_ems_paper}.

\end{abstract}


\begin{teaserfigure}
  \includegraphics[width=\columnwidth]{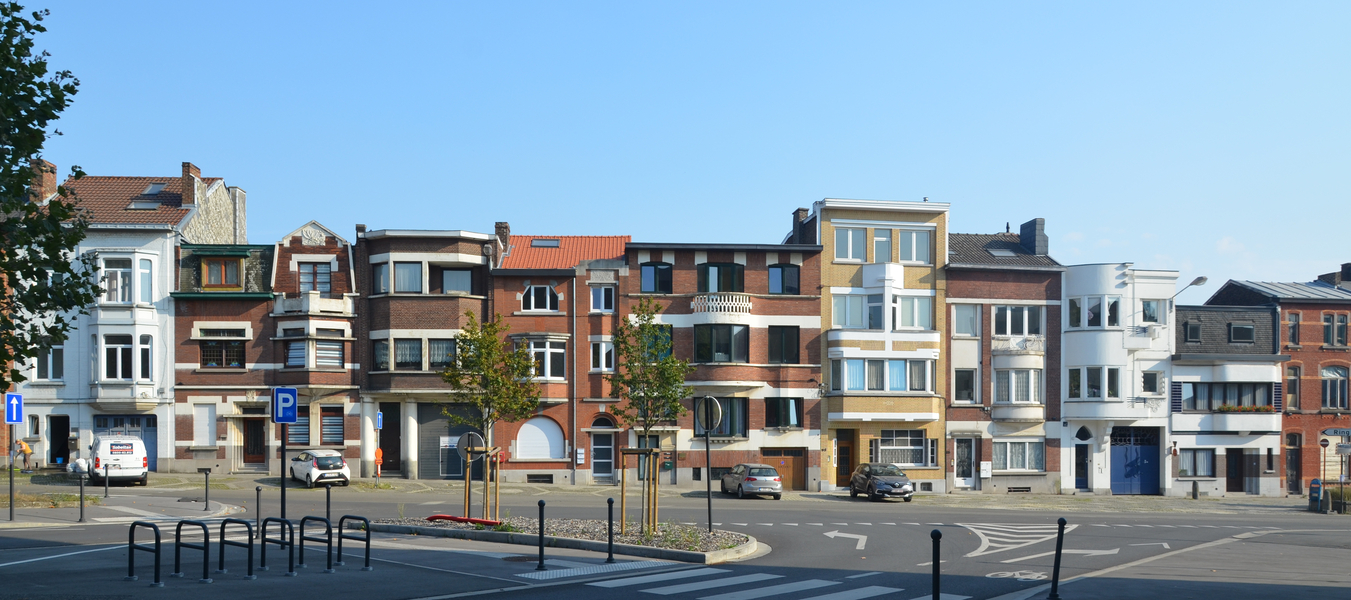}
  \caption{1 Rue de la Paix, Ville-Basse, Charleroi, Belgium \cite{jmh2oFrancaisCharleroiBelgique2020}.}
  \Description{A residential street in Charleroi, Belgium, showing a mix of dwellings.}
  \label{fig:teaser}
\end{teaserfigure}

\maketitle

\section{Introduction}

An effective management of energy assets in residential buildings is becoming increasingly relevant in Belgium.
This is driven by the increasing electrification of our energy systems \cite{vbofebBelgianMobilityDashboard2025,vlaamsenergie-enklimaatagentschapZonnepanelenVlaanderen2024}, the possibility to adopt hourly electricity prices, and the implementation of grid peak tariffs \cite{vlaamsenutsregulatorCapaciteitstarief2025,ruddickRealworldValidationSafe2024}.
As these systems and regulations become more complex, the question arises: What are the benefits of implementing an advanced energy management system in electrified dwellings ? 

This study aims to answer this question by simulating a set of representative Belgian dwellings which have been electrified and are subject to dynamic pricing and grid peak tariffs.
The electrification of the dwellings is achieved by replacing the traditional heating systems with air-to-air heat pumps, adding domestic electric vehicle (EV) chargers due to the replacement of petrol cars with EVs, and replacing gas water heaters with electric water heaters.
Other dwelling characteristics such as the presence of photovoltaic (PV) panels, insulation level, and construction type are kept close to current Belgium situation according to the most recent statistics we could find.
Real-world datasets are used for base electricity consumption, domestic water usage, EV charging sessions, PV production, weather data, and electricity tariffs. Two years of data (2022 for training, 2023 for evaluation) are used, reflecting a realistic scenario where a household adopts an energy management system (EMS) in January 2023 based on data from the previous year.
Home batteries are excluded from the analysis, as their adoption remains limited in Belgium. According to available statistics, about 3\% of dwellings had a home battery in 2024 \cite{vangastelHardeCijfersVorig2025}. Replacing gas boilers with heat pumps and petrol cars with EVs when necessary is essential for reducing fossil fuel use, while the case for home batteries is less evident.

This study compares three distinct EMS strategies: rule-based control (RBC), which mimics the default operation of residential energy assets; model predictive control (MPC), a state-of-the-art optimization-based approach; and TreeC, a machine learning-based method that models EMS strategy as decision trees.
In addition to these realistic EMSs, a model predictive control EMS with perfect forecast (MPC P) is also compared to show a maximum performance an EMS could technically achieve.

Buying a home EMS from a professional installer is advertised as costing between 280\,€ and 680\,€ with some additional subscription costs ranging between 0\,€ and 360\,€ per year \cite{maakjemeterslimAlleEMSProducten2025}.
 An EMS can be implemented cheaper using for example Home Assistant and the additional EMHASS module \cite{homeassistantHomeAssistant2025,davidusb-geekEnergyManagementHome2025} for which the basic hardware costs close to 100\,€ but requires some technical knowledge to set up and configure.
To these basic costs can be added the maintenance, expensive compatible equipment and additional necessary sensors.

This study addresses the following research questions for a representative set of electrified Belgian dwellings:

\begin{itemize}
    \item What are the benefits of EMSs like TreeC and MPC compared to a simple RBC in terms of cost savings, thermal comfort and EV charging satisfaction ?
    \item How much can algorithmic improvements realistically improve these benefits ?
    \item What is the influence of different dwelling characteristics on cost savings ?
\end{itemize}

\section{Case description: set of 500 representative electrified Belgian dwellings}

This section resumes the important aspects of the 500 representative electrified Belgian dwellings which is further detailed in section~\ref{sec:representative_dwellings} of the appendix\footnote{Appendix: \url{https://raw.githubusercontent.com/EVERGi/should_belgians_install_advanced_ems_paper/refs/heads/main/article_with_appendix.pdf\#page=12}\label{foot:appendix}} %
.
The electricity cost used in this study is extensively described in this section, as it is a key component to understand the results and can be complex for readers not familiar with the Belgian context.

\subsection{Dwelling generation, data sources and simulation models}

Each dwelling is first assigned a set of characteristics (such as region, number of rooms, and insulation level) based on the latest available Belgian statistics (see section~\ref{sec:dwelling_characteristics}). Using these characteristics, a simulation model is created for each dwelling, incorporating real-world data sources (see \cref{sec:thermal_data_and_models,sec:consumption_profiles,sec:ev_charger,sec:pv_panels,sec:hot_water}). This process yields 500 unique simulation models, each reflecting its own set of characteristics and data inputs. The overall methodology is illustrated by the flow chart in \cref{fig:dwelling_flow_chart}.

Of the 500 simulated electrified dwellings, 33.8\% are apartments and 66.2\% are houses. The average surface area is 89.4\,m\textsuperscript{2} (standard deviation 35.7\,m\textsuperscript{2}) \cite{vannesteLogementBelgique2007}. In total, 20.6\% have no insulation, 48.6\% have partial insulation (roof, wall and/or ground), and 30.8\% have full insulation (roof, wall and ground)\cite{recticelinsulationBarometreLisolation20202020}. Moreover, 68.9\% are assigned an EV charger \cite{hubertMobiliteQuotidienneBelges2003}, 25.8\% have PV panels\cite{brugelEnergiesRenouvelablesRegion2024,vlaamsenergie-enklimaatagentschapZonnepanelenVlaanderen2024,lallemand16MenagesWallons2023}, and 10.6\% have no central heating so no heat pump and no electric water heater \cite{statbelCensus2021Nombre2021}.

Most characteristics are region-specific. For example, among dwellings located in Brussels, a larger share are apartments than in the other regions. This applies to all characteristics except surface area and number of bedrooms, which depend on the number of rooms in the dwelling instead \cite{vannesteLogementBelgique2007,meirelesDomesticHotWater2022}. Hot water consumption is also correlated with the number of bedrooms \cite{meirelesDomesticHotWater2022}. Apart from this, the characteristics and real-world data are sampled independently, since no data on additional correlations was found in the used refferences.

The whole simulation methodology, including datasets used and building models, is described in detail in section~\ref{sec:representative_dwellings} of the appendix\textsuperscript{\ref{foot:appendix}}. %

The authors acknowledge limitations of the methodology, including the use of EV charging behaviour data from Norway \cite{sorensenElectricVehicleCharging2024} and the dwelling surface area being independent of PV system size. These limitations arise because credible sources to quantify specific correlations between characteristics were not found, or because no openly accessible dataset was available for the Belgian context. Nevertheless, the methodology relies on the best available data to our knowledge and is designed to be transparent and reproducible, enabling future improvements as additional data become available.

\newcommand{\hconnect}[2]{%
    \draw[thick,->] (#1.east) -- ++(0.564,0) |- (#2.west);
}
\newcommand{\vconnect}[2]{%
    \draw[thick,->] ([xshift=-1.3cm]#1.south) |- (#2.west);
}
\newcommand{\hshortconnect}[2]{%
    \draw[thick,->] (#1.east) -- ++(0.15,0) |- (#2.west);
}
\newcommand{\hlongconnect}[2]{%
    \draw[thick,->] (#1.east) -- ++(6.864,0) |- (#2.west);
}

\newlength{\defaultwidth}
\setlength{\defaultwidth}{2.9cm}

\newlength{\xshiftsubcategory}
\setlength{\xshiftsubcategory}{\dimexpr\defaultwidth/7\relax}

\newlength{\middlewidth}
\setlength{\middlewidth}{3.5cm}

\newlength{\rightwidth}
\setlength{\rightwidth}{2.6cm}

\newlength{\pvoutputwidth}
\setlength{\pvoutputwidth}{\dimexpr\middlewidth-1cm\relax}

\newlength{\hpxmlwidth}
\setlength{\hpxmlwidth}{\dimexpr\middlewidth-0.7cm\relax}

\newlength{\energypluswidth}
\setlength{\energypluswidth}{\dimexpr\middlewidth/2\relax}

\newlength{\belowleftrow}
\setlength{\belowleftrow}{0.2cm}

\aptLtoX[graphic=no,type=html]{\begin{figure*}
\centering
\includegraphics[width=.9\textwidth]{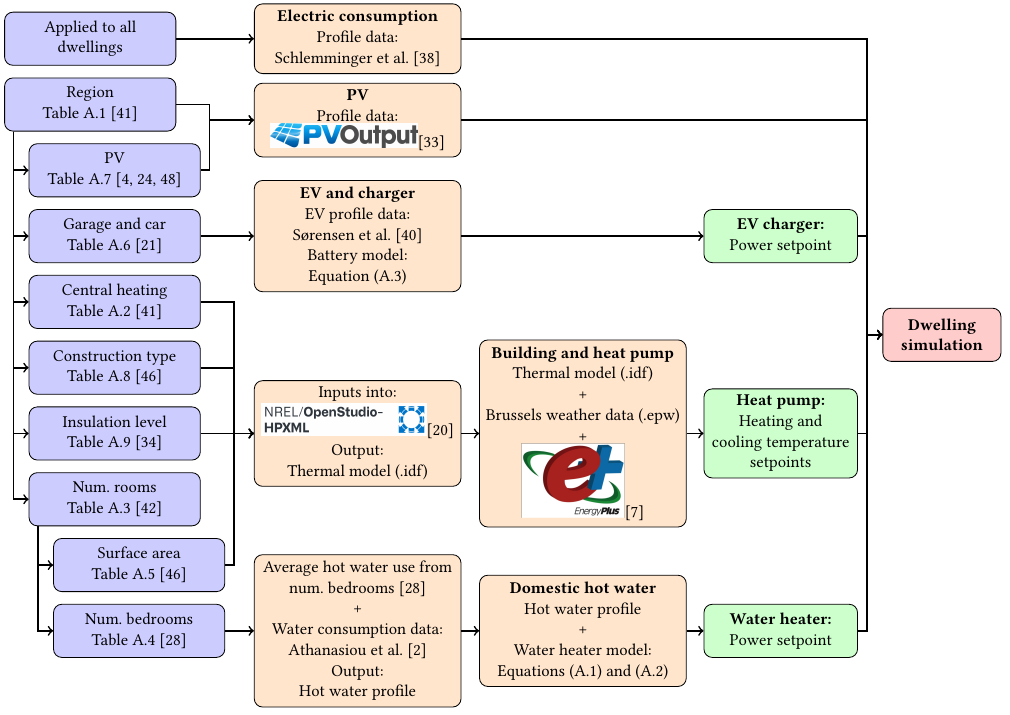}
    \caption{Flow chart of this study's methodology to simulate a dwelling. Blue nodes represent the sampled characteristics of the dwelling with associated tables and sources (section~\ref{sec:dwelling_characteristics}), orange nodes represent the data and simulation models used for each part of the dwelling (\cref{sec:thermal_data_and_models,sec:consumption_profiles,sec:ev_charger,sec:pv_panels,sec:hot_water}), green nodes represent the possible control actions of the assets (\cref{sec:asset_management}), and the red node is the dwelling simulation that combines everything together.}
    \label{fig:dwelling_flow_chart}
    \Description{Flow chart illustrating the methodology for simulating a dwelling. The chart is divided into four main sections: dwelling characteristics (blue nodes), data and simulation models (orange nodes), asset management (green nodes), and the dwelling simulation (red node). Each section contains specific components, such as region, PV, garage, central heating, construction type, insulation level, number of rooms, surface area, number of bedrooms, electric consumption profile, PV profile, EV and charger profile, thermal and heat pump model, domestic hot water model, and control actions for EV charger, heat pump, and water heater. Arrows indicate the flow of information and connections between these components, demonstrating how they interact to create a comprehensive dwelling simulation.}
\end{figure*}}{
\begin{figure*}
    \centering
    \begin{tikzpicture}[
        node distance=0.8cm,
        every node/.style={draw, rounded corners, font=\small, align=center, fill=blue!20, minimum width=\defaultwidth, minimum height=0.9cm},
        every edge/.style={draw,->,thick},
    ]
    \node (all) at (0,0) {Applied to all \\dwellings};
    \node[below=\belowleftrow of all] (region) {Region \\ 
    Table~\ref{tab:stats_region} \cite{statbelCensus2021Nombre2021}};
    \node[below=\belowleftrow of region, xshift=\xshiftsubcategory] (pv) {PV \\ Table~\ref{tab:stats_pv} \cite{brugelEnergiesRenouvelablesRegion2024,vlaamsenergie-enklimaatagentschapZonnepanelenVlaanderen2024,lallemand16MenagesWallons2023}};
    \node[below=\belowleftrow of pv] (garage) {Garage and car \\ Table~\ref{tab:stats_home_charger} \cite{hubertMobiliteQuotidienneBelges2003}};
    \node[below=\belowleftrow of garage] (central) {Central heating \\ Table~\ref{tab:stats_heating} \cite{statbelCensus2021Nombre2021}};
    \node[below=\belowleftrow of central] (construction) {Construction type \\ Table~\ref{tab:stats_construction_type} \cite{vannesteLogementBelgique2007}};
    \node[below=\belowleftrow of construction] (insulation) {Insulation level \\ Table~\ref{tab:stats_insulation} \cite{recticelinsulationBarometreLisolation20202020}};
    \node[below=\belowleftrow of insulation] (rooms) {Num. rooms \\ Table~\ref{tab:stats_rooms} \cite{statbelCensus2021Nombre2021a}};

    \node[below=\belowleftrow of rooms, xshift=\xshiftsubcategory] (surface) {Surface area \\ Table~\ref{tab:stats_surface_area} \cite{vannesteLogementBelgique2007}};
    \node[below=\belowleftrow of surface] (bedrooms) {Num. bedrooms \\ Table~\ref{tab:stats_bedrooms} \cite{meirelesDomesticHotWater2022}};

    \node[right=1.315cm of all, minimum width=\middlewidth, fill=orange!20] (elecprof) {\textbf{Electric consumption}\\
    Profile data: \\ 
    \citet{schlemmingerDatasetElectricalSinglefamily2022}};
    \node[right=0.9cm of pv,yshift=0.85cm, minimum width=\middlewidth, fill=orange!20] (pvprof) {\textbf{PV} \\ Profile data: \\
      \includegraphics[width=\pvoutputwidth
      ]{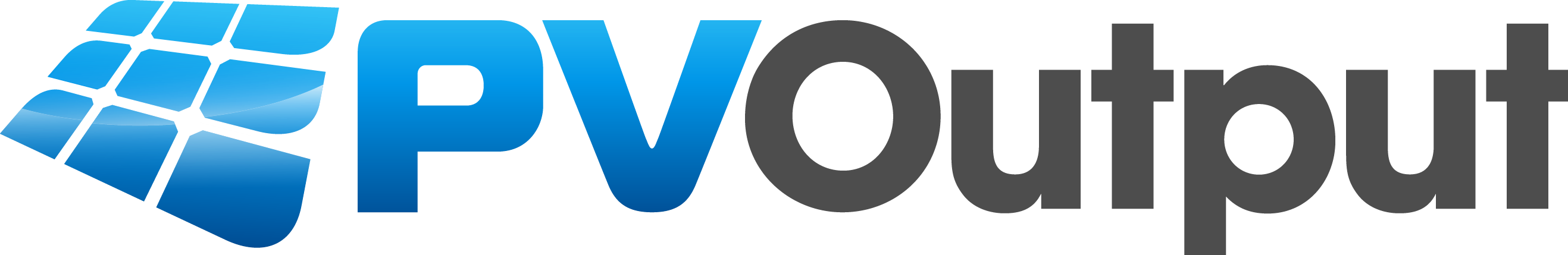}\cite{pvoutput.orgPVOutputorg2025}};
    \node[right=0.9cm of garage, minimum width=\middlewidth, fill=orange!20] (evprof) {\textbf{EV and charger} \\ 
    EV profile data: 
    \\ \citet{sorensenElectricVehicleCharging2024} \\
    Battery model: \\ Equation~\eqref{eq:ev_charging_reliability}};

    \node[below=1.492cm of evprof, minimum width=\middlewidth, fill=orange!20] (hpxml) {Inputs into: \\\includegraphics[width=\hpxmlwidth]{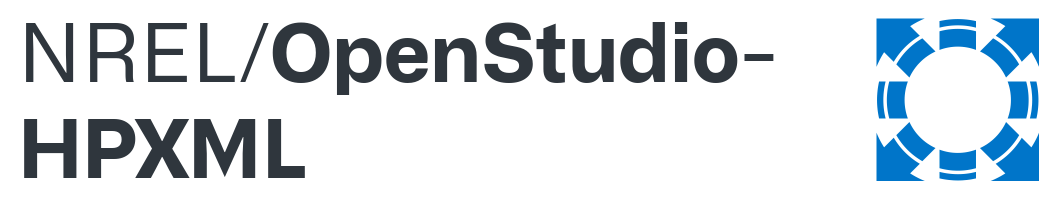}\cite{horowitzOpenStudioHPXMLWorkflow2024}\\ Output: \\Thermal model (.idf)};

    \node[right=0.3cm of hpxml, minimum width=\middlewidth, fill=orange!20] (heatmodel) {\textbf{Building and heat pump}\\
    Thermal model (.idf)\\
    +\\
    Brussels weather data (.epw) \\
    +\\
    \includegraphics[width=\energypluswidth]{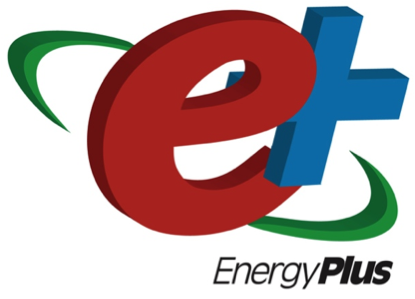}\cite{crawleyEnergyPlusCreatingNewgeneration2001}
    };

    \node[below=1.144cm of hpxml, minimum width=\middlewidth, fill=orange!20] (hot_water) {Average hot water use from \\num. bedrooms \cite{meirelesDomesticHotWater2022}
     \\
    +\\
    Water consumption data:\\
    \citet{athanasiouDAIADOpenWater2014}\\ Output: \\
    Hot water profile};
    \node[right=0.3cm of hot_water, minimum width=\middlewidth, fill=orange!20] (dhwmodel) {\textbf{Domestic hot water}\\
    Hot water profile\\
    +\\
    Water heater model:\\ Equations~(\ref{eq:hot_water_params}) and (\ref{eq:tank_temp_change})};
    
    \node[right=4.1cm of evprof, minimum width=\rightwidth, fill=green!20] (evcontrol) {\textbf{EV charger:}\\
    Power setpoint};
    \node[below=2.123cm of evcontrol, minimum width=\rightwidth, fill=green!20] (heatcontrol) {\textbf{Heat pump:}
    \\ Heating and \\cooling 
    temperature \\setpoints};
    \node[below=2.125cm of heatcontrol, minimum width=\rightwidth, fill=green!20] (dhwcontrol) {\textbf{Water heater:}\\ Power setpoint};

    \path (construction) -- (central) coordinate[midway] (midctrl);

    \node[right=13cm of midctrl,yshift=0cm, minimum width=2cm, fill=red!20] (dwsim) {\textbf{Dwelling}\\ \textbf{simulation}};

    \hconnect{all}{elecprof}
    \hconnect{region}{pvprof}
    \vconnect{region}{pv}
    \vconnect{region}{garage}
    \vconnect{region}{central}
    \vconnect{region}{construction}
    \vconnect{region}{insulation}
    \vconnect{region}{rooms}

    \hshortconnect{pv}{pvprof}
    \hconnect{garage}{evprof}
    \hconnect{central}{hpxml}
    \hconnect{construction}{hpxml}
    \hconnect{insulation}{hpxml}
    \vconnect{rooms}{surface}
    \hshortconnect{surface}{hpxml}
    \draw[thick, ->] (hpxml) -- (heatmodel);
    
    \vconnect{rooms}{bedrooms}
    \hshortconnect{bedrooms}{hot_water}
    \draw[thick, ->] (hot_water) -- (dhwmodel);

    \hlongconnect{elecprof}{dwsim}
    \hlongconnect{pvprof}{dwsim}
    \hshortconnect{evprof}{evcontrol}
    \hshortconnect{heatmodel}{heatcontrol}
    \hshortconnect{dhwmodel}{dhwcontrol}

    \hshortconnect{evcontrol}{dwsim}
    \hshortconnect{heatcontrol}{dwsim}
    \hshortconnect{dhwcontrol}{dwsim}

    \end{tikzpicture}
    \caption{Flow chart of this study's methodology to simulate a dwelling. Blue nodes represent the sampled characteristics of the dwelling with associated tables and sources (section~\ref{sec:dwelling_characteristics}), orange nodes represent the data and simulation models used for each part of the dwelling (\cref{sec:thermal_data_and_models,sec:consumption_profiles,sec:ev_charger,sec:pv_panels,sec:hot_water}), green nodes represent the possible control actions of the assets (\cref{sec:asset_management}), and the red node is the dwelling simulation that combines everything together.}
    \label{fig:dwelling_flow_chart}
    \Description{Flow chart illustrating the methodology for simulating a dwelling. The chart is divided into four main sections: dwelling characteristics (blue nodes), data and simulation models (orange nodes), asset management (green nodes), and the dwelling simulation (red node). Each section contains specific components, such as region, PV, garage, central heating, construction type, insulation level, number of rooms, surface area, number of bedrooms, electric consumption profile, PV profile, EV and charger profile, thermal and heat pump model, domestic hot water model, and control actions for EV charger, heat pump, and water heater. Arrows indicate the flow of information and connections between these components, demonstrating how they interact to create a comprehensive dwelling simulation.}
\end{figure*}}

\subsection{Electricity cost}

The real-time pricing used in this study is based on a one-year dynamic contract of a major Belgian energy supplier, as if it were taken in January 2023\footnote{Full contract: 
\url{https://github.com/EVERGi/should_belgians_install_advanced_ems_paper}}.
The total cost is composed of four costs with different pricing mechanisms: the day-ahead cost, the offtake extras cost, the peak cost and the fixed yearly cost.

The day-ahead cost depends on the prices of \cref{eq:injection_price_usefull,eq:offtake_price_usefull}.

\begin{equation}
    Vo_t= 
\begin{cases}
    (Vd_t+0.002\frac{\text{€}}{kWh}) \times 1.06& \text{if } (Vd_t+0.002\frac{\text{€}}{kWh})\geq 0\\
    (Vd_t+0.002\frac{\text{€}}{kWh}),              & \text{otherwise}
\end{cases}
\label{eq:injection_price_usefull}
\end{equation}

\begin{equation}
    Vi_t=Vd_t\frac{\text{€}}{kWh}
\label{eq:offtake_price_usefull}
\end{equation}

where:

\begin{tabular}{ll}
    $Vo_t$ [€/kWh] & Offtake price at time step $t$ \\
    $Vi_t$ [€/kWh] & Injection price at time step $t$ \\
    $Vd_t$ [€/kWh] & Belgian day-ahead hourly price  at time step $t$ \\
\end{tabular}

A value-added tax of 6\% is added to the offtake price when it is positive.
Extra costs that occur for energy taken from the grid for transport, distribution and taxes are calculated together and labelled as offtake extras cost.
The offtake price for all these additional offtake costs $Vf$ is 0.079 €/kWh. 

The monthly peak offtake cost $Cp_m$ is calculated by \cref{eq:peak_cost_usefull}.
\begin{equation}
Cp_m = Vp \times \max_{{t\in B_m}} (\frac{Eo_t}{\Delta t}, 2.5 kW)
\label{eq:peak_cost_usefull}
\end{equation}

where:

\begin{tabular}{ll}
    $Vp$ [€/kW] & Peak price (used: 3.66 €/kW) \\
    $Eo_t$ [kWh] & Energy taken from the grid at time step $t$ \\
    $B_m$ & Set of time steps in month $m$ \\
    $\Delta t$ [h] & Length of time step \\
\end{tabular}

This cost adds a penalty to the highest offtake power measured over any 15 minute time step within each month. 
The offtake energy is divided by the length of a time step to obtain the power in kW.
A minimum monthly peak power of 2.5 kW is enforced in the cost, even if the measured peak power is lower.

The monthly peak offtake cost, otherwise called capacity tariff, is a pricing mechanism applied to all residential consumers equipped with a digital meter in the Belgian region of Flanders.
The goal is to incentivize consumers to lower their peak offtake power from the grid.

Finally, part of the electric bill is a fixed yearly cost $Cy$ of 115.23\,€ per year.

The total electricity cost $Ctot$ is calculated by \cref{eq:tot_cost_usefull}.

\begin{equation}
Ctot=\underbrace{\sum_{t\in T}{(Eo_tVo_t-Ei_tVi_t)}}_{\text{day-ahead}}+\underbrace{\sum_{t\in T}Eo_tVf}_{\text{offtake extras}}+\underbrace{\sum_{m \in M}{Cp_m}}_{\text{peak}}+\underbrace{Cy}_{\text{yearly}}
\label{eq:tot_cost_usefull}
\end{equation}

where:

\begin{tabular}{ll}
    $T$ & Set of all time steps \\
    $M$ & Set of all months \\
\end{tabular}

\section{Asset management}
\label{sec:asset_management}

The three controllable assets, when present in the dwelling, are the air-to-air heat pump, the water heater and the EV charger.

Every simulation is run for a period of one year with a time step of 15 minutes as this is the time period used to calculate the peak electricity cost in Belgium (see \cref{eq:peak_cost_usefull}).
The EMS can change at each time step the cooling and heating temperature set points of the heat pump, the electric power of the water heater and the charge power of the EV charger.

\subsection{Heat pump control}
\label{sec:heat_pump_control}

The heat pump operates within a default comfort range.
If the indoor temperature is outside the comfort range, the heat pump will bring the indoor temperature back to the comfort range. 
The EMS can set a heating or cooling temperature set point within this comfort range.

Based on different studies, \citet{peetersThermalComfortResidential2009} established usable comfort ranges for the different rooms of a residential building.

The comfort range depends on the reference external temperature $T^{e,ref}_d$ which is based on the minimum and maximum temperatures of the current and the last three days as shown in \cref{eq:te_ref}.

\begin{equation}
    T^{e,ref}_d = \frac{T^{avg}_{d} + 0.8 \times T^{avg}_{d-1} + 0.4 \times T^{avg}_{d-2} + 0.2 \times T^{avg}_{d-3}}{2.4}
\label{eq:te_ref}
\end{equation}

where:

\begin{tabular}{ll}
    $T^{avg}_{d}$ & \begin{tabular}{@{}l@{}}Average of the maximum and minimum temperature \\ on day $d$ \end{tabular} \\
\end{tabular}

To avoid relying on weather forecasts for $T^{avg}_{d}$, this study uses $T^{avg}_{d-1}$ as a substitute in the calculation. The mean absolute difference in the resulting comfort range is only 0.18\,°C (standard deviation 0.24\,°C), showing that this simplification has minimal impact while reducing forecast dependency.

\begin{figure}
\centering
\includegraphics[width=\columnwidth]{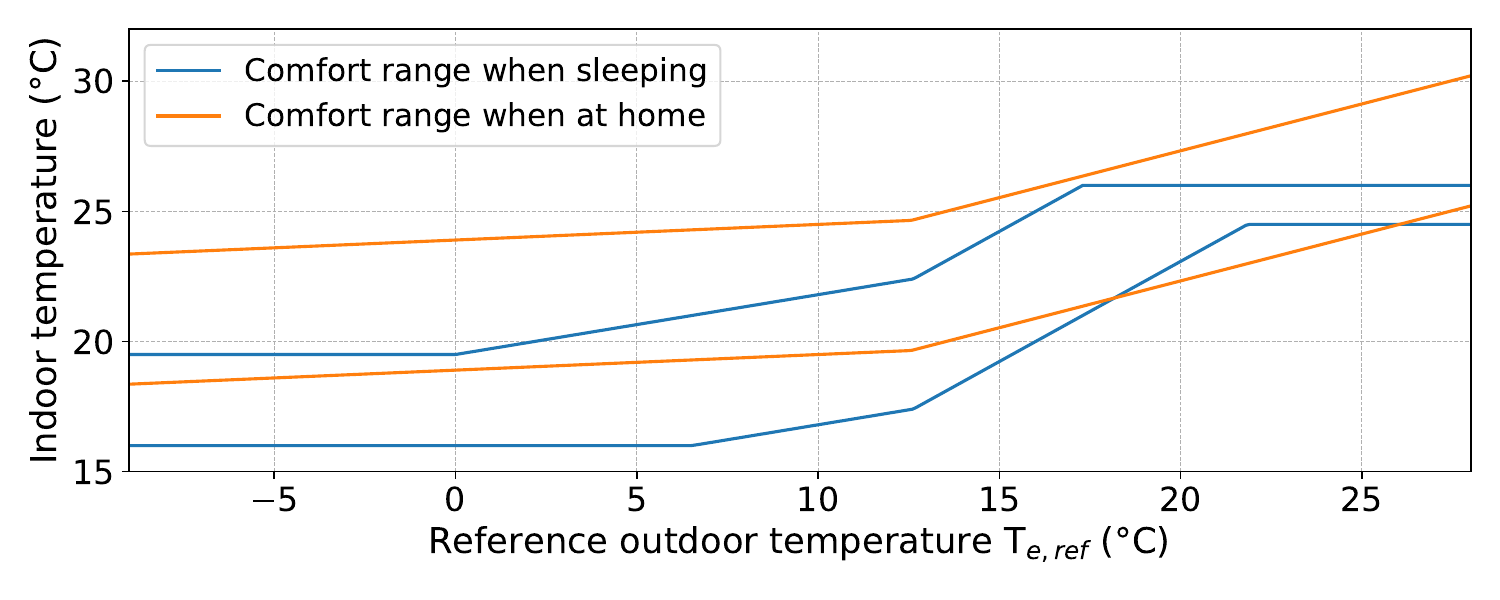}
\caption{
    Illustration of the comfort temperature ranges for sleeping and home occupancy profiles as a function of the reference external temperature $T^{e,ref}_d$. The plot covers $T^{e,ref}_d$ values from $-9\,^\circ$C to $28\,^\circ$C, representing the coldest and warmest reference temperatures observed in Belgium during the 21st century.
}
\Description{A line graph showing the comfort temperature ranges for sleeping and home occupancy profiles as a function of the reference external temperature T {e,ref} d. The x-axis represents the reference external temperature ranging from -9°C to 28°C, while the y-axis represents the comfort temperature in °C. Two lines are plotted: one for sleeping comfort range and another for home occupancy comfort range. The sleeping comfort range is lower than the home occupancy comfort range across all reference external temperatures.}
\label{fig:comfort_range}
\end{figure}

Based on $T^{e,ref}_d$, comfort ranges are defined using a width $w$ = 5\,°C and asymmetry $\alpha$ = 0.7, corresponding to 10\% predicted percentage of dissatisfied users \cite{peetersThermalComfortResidential2009}.
Two different comfort ranges are used: sleeping and at home.
\Cref{fig:comfort_range} illustrates the comfort ranges for both profiles as a function of the reference external temperature $T^{e,ref}_d$. For an example of how the comfort range evolve throughout a weekend, check \cref{fig:example_weekend}.

\citet{aertsDiscreteOccupancyProfiles2013} defines a set of occupancy profiles out of 3\,474 Belgian households.
The paper finds 7 daily occupancy profiles representative of people's behaviour which cover 79\% of the evaluated households.
The occurrence of each profile depends on whether the day is a weekday, Saturday, or Sunday.
In these profiles, the household is either at home, absent or sleeping for each time step of the day.

Each dwelling is assigned one of the 7 representative occupancy profiles for the weekdays, Saturdays, and Sundays based on the distribution provided by \citet{aertsDiscreteOccupancyProfiles2013}.
The occupancy profiles are used to determine the comfort range based on whether the household is at home, sleeping, or absent at each time step.

\subsection{Water heater control}

The hot water tank has a temperature range from 55 °C to 65 °C that it should maintain.
These values are the minimum and maximum domestic hot water temperatures from the European, Australian, Canada, Japan and Unites States standards \cite{fuentesReviewDomesticHot2018}. 
If the water temperature goes below 55 °C, the water heater will heat the water at maximum power for one 15 minute time step. Whenever the water temperature is above 65 °C, the water heater will not heat the water.
Otherwise, the water can be heated between 0 kW and the maximum power of the water heater.

\subsection{Electric vehicle charger control}
\label{sec:ev_charger_control}

The EV charger is controlled by setting a power between 0 kW and 3.7 kW.
Once the required departure state of charge is reached, charging stops.
Currently, most EVs do not provide information about their battery's state of charge \cite{sorensenMethodGeneratingComplete2023}. Therefore, in this work, the EMSs are assumed to operate without knowledge of the battery's state of charge.

For dwellings with an EV charger, two types of simulations are run. In the first, the EMS fully controls EV charging, allowing to assess how well each EMS charges the EV (see \cref{fig:performance_comparison}d). In the second, charging is enforced so the EV always reaches full charge before departure, ensuring a fair cost comparison across EMSs (see %
figs.~\ref{fig:performance_comparison}a, \ref{fig:performance_comparison}b, \ref{fig:dwelling_features_results} and \ref{fig:opex_diff_separate}).

In the enforced scenario, if the EV is not fully charged near departure, the charger switches to 3.7 kW to guarantee a full charge, simulating a driver’s request for a complete charge before leaving.

\section{Energy management systems}

\subsection{Rule-based control}
\label{sec:relevance_rbc}
The rule-based control EMS uses simple, predefined strategies for each controllable asset, reflecting typical default operation modes of these assets:

\begin{itemize}
    \item \textbf{Heat pump:} The heat pump setpoint is determined by obtaining the comfort range for the next 1 hour and 30 minutes. The controller gets the highest lower bound ($T^{low}_t$) and the lowest upper bound ($T^{up}_t$) during this period and sets the heating and cooling setpoints to these values.
    \item \textbf{Domestic water heater:} The water heater operates at maximum power for 15 minutes whenever the tank temperature drops below 55\,°C.
    \item \textbf{Electric vehicle charger:} The EV is always charged at the maximum power.
\end{itemize}

The 1 hour and 30 minute time horizon was selected based on the final results to ensure that the rule-based controller achieves a thermal discomfort level similar to the other EMSs (see \cref{fig:performance_comparison}c).

\subsection{Model predictive control}

Model predictive control (MPC) is a control strategy that optimizes the operation of controllable assets over a prediction horizon by solving an optimization problem at each time step.
The MPC uses a mathematical model of the assets and predicts all necessary parameters such as weather, electricity prices, and energy consumption of non-controllable assets to define the set points for the controllable assets.

This study implements two MPC approaches: a realistic MPC (MPC) and a perfect forecast MPC (MPC P).
The realistic MPC is a state-of-the-art control algorithm for EMSs. The perfect forecast MPC serves as an unrealistically performant EMS to indicate an upper EMS performance limit.

The realistic MPC uses realistic forecasts for most parameters and uses a prediction horizon that extends up to when day-ahead prices ($Vd_t$) are available, that is until midnight of the current day if before 14:00 Central European Time or until midnight of the following day if after 14:00, resulting in a horizon ranging from 10 hours and 15 minutes to 34 hours.
MPC P uses a perfect forecast for all parameters and has a fixed prediction horizon of 48 hours.

Both MPC approaches are implemented using the Gurobi solver \cite{gurobioptimizationllcGurobiOptimizerReference2024}.

\subsubsection{Mathematical model and objective function}
The MPC's model is formulated as a mixed-integer linear program (MILP). The thermal model of the dwellings is approximated using the resistor-capacitor analogy for thermal modelling \cite{madsenEstimationContinuoustimeModels1995,drgonaAllYouNeed2020}. The water heater and EV charging are modelled using the same equations as the simulation. The constraints aim to not violate comfort ranges and to charge the EV before departure. The objective function minimizes the electricity cost over the prediction horizon.
A full description of the MPC's mathematical model is available in the appendix\textsuperscript{\ref{foot:appendix}} (see section~\ref{sec:mpc_model}).

\subsubsection{Forecast}

The forecasted variables used by the MPC's model are:

\begin{itemize}
    \item The outdoor temperature ($T^{out}_{t}$)
    \item The solar irradiance ($Q^{sun}_{t}$)
    \item The hot water flow ($F_t$)
    \item The arrival time of the EV
    \item The departure time of the EV
    \item The charged energy of the EV
    \item The PV production ($P^{PV}_{t}$)
    \item The base load consumption ($P^{load}_{t}$)
\end{itemize}

MPC P uses the actual future values of all relevant variables over its 48-hour prediction horizon.

For the realistic MPC, all non EV related variables in the next hour are predicted using the most recent value, while subsequent time steps are predicted using values from the preceding 24 hours at the same time of the day.

Predicting the arrival time of an EV is particularly challenging due to the high variability in user behaviour.
Therefore, the MPC only assumes that an EV arrives when a new charging session starts. For each new charging session, the departure time and charged energy are predicted using k-Nearest Neighbor (k-NN) \cite{coverNearestNeighborPattern1967} regression as it showed good performance in \citet{genovForecastingFlexibilityCharging2024} to predict departure time and charged energy of EVs.
The forecasting features used are the arrival time and week day, the charged energy from the previous charging session, the total connection time of the previous charging session, and the time since the start of the previous charging session. The model is trained on the 2022 charging profile data of the dwelling the MPC is controlling and the k-NN model is retrained with one additional charging session data every time a new charging session is finished during simulation.

To account for prediction errors, if the MPC incorrectly predicts that the EV is fully charged but it remains connected and not fully charged, the MPC will schedule charging at 3.7 kW for the next 2 hours to ensure additional charging. Conversely, if the MPC overestimates the departure time and the enforced charging behaviour was triggered at the previous time step (see \cref{sec:ev_charger_control}), then the MPC assumes the EV will be charged at maximum power for the next 2 hours.

\subsection{TreeC}

TreeC is a machine learning \footnote{Justification for machine learning classification in Chapters 2 and 4 of \citet{ruddickHowEffectivelyManage2026}.} based energy management system that uses a metaheuristic algorithm to generate decision trees that define a control strategy. 
It evaluates various decision tree strategies on the dwelling's simulation and selects the best performing one. This yields a tailored and interpretable control policy for each dwelling \cite{ruddickTreeCMethodGenerate2025}.

\subsubsection{Training process and objective function}

The training evaluates different decision tree based EMSs on the 2022 simulation. A particle swarm optimization (PSO) algorithm \cite{kennedyParticleSwarmOptimization1995} generates and optimizes these trees by selecting splitting features, thresholds, and setpoints for each asset. The trees are pruned to maintain interpretability and reduce overfitting.

The objective function minimizes electricity costs, maintains indoor comfort, and ensures sufficient EV charging. 

A full description of the training and objective function is available in the appendix\textsuperscript{\ref{foot:appendix}} (see section~\ref{sec:treec_objective_function}).

\subsubsection{Decision tree model}

The training can select from up to 23 different inputs to use as splitting features in the decision trees.
The inputs related to specific assets are not used in the dwellings where the asset is not present.

The inputs are:
\begin{itemize}
\item The time of the day
\item The day of the week
\item The cosine of the day of the year, normalized so that $-1$ corresponds to December 21\textsuperscript{st} (winter solstice) and $+1$ to June 21\textsuperscript{st} (summer solstice)
\item $P^{asset}_{t-1}$, the average power of each asset (heat pump, water heater, the EV charger, base load, PV) and total power for the previous time step (6 inputs)
\item $T^{in}_{t}$, the indoor temperature
\item $T^{low}_{t}$, the lower bound of the comfort range for the next time step, in 30 minutes and in 2 hours (3 inputs)
\item $T^{tank}_{t}$, the water tank temperature
\item $T^{out}_{t}$, the outdoor temperature
\item $Q^{sun}_{t}$, the solar irradiance
\item $Vd_t$, the day-ahead price
\item $\max_{{h\in H}}(Vd_h)-Vd_t$ the difference between the maximum and the current price for the next 6, 12 and 24 hours with $H$ being the set of future time steps (3 inputs)
\item $Vd_t-\min_{{h\in H}}(Vd_h)$, the difference between the current price and the minimum price for the next 6, 12 and 24 hours with $H$ being the set of future time steps (3 inputs)
\end{itemize}

The TreeC EMS can control the heat pump through three types of setpoints: comfort rule-based, temperature setpoint and adjustment to indoor temperature.
The comfort rule-based setpoint in TreeC follows the same logic as in the rule-based controller: it sets the heat pump setpoints for cooling and heating to match the comfort range over a specified time window (see \cref{sec:relevance_rbc}). Unlike the rule-based controller, however, TreeC select the length of this time window anywhere from 0 to 12 hours.
The temperature setpoint sets the heat pump setpoints for cooling and heating to a fixed temperature between 5,\,°C and 40\,°C. 
The adjustment to indoor temperature setpoint sets the heat pump's heating and cooling setpoints relative to the current indoor temperature $T^{in}_{t}$. For heating, the setpoint is defined as $T^{in}_{t}$ plus a fixed offset; for cooling, as $T^{in}_{t}$ minus a fixed offset. This fixed offset can be selected within the range of $-2\,^\circ$C to $5\,^\circ$C.

The TreeC EMS control the water heater and EV charger by setting directly the power of the asset.

\subsubsection{Interpretable decision tree}

\begin{figure}
    \centering
    \begin{subfigure}{0.495\columnwidth}
        \includegraphics[width=\linewidth]{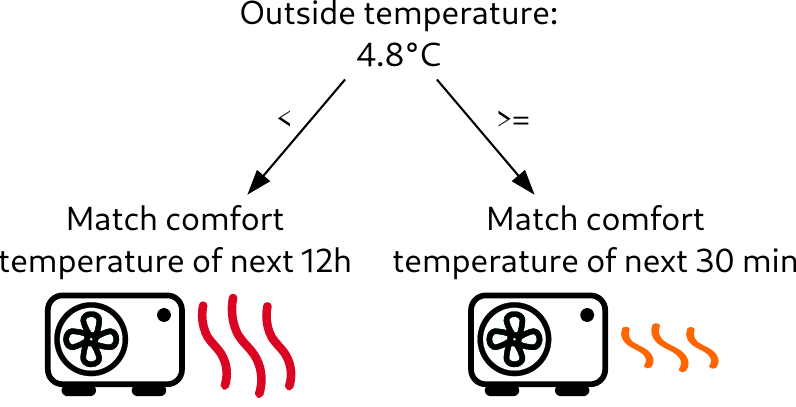}
        \caption{Heating setpoint
        }
        \label{fig:heating_decision_tree}
    \end{subfigure}
    \hfill
    \begin{subfigure}{0.495\columnwidth}
        \includegraphics[width=\linewidth]{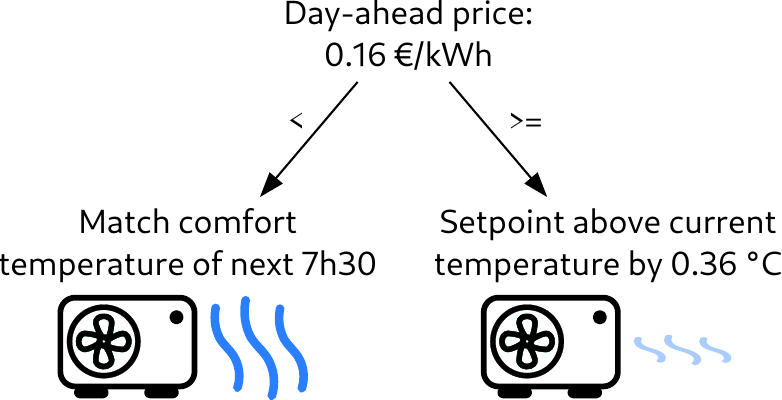}
        \caption{Cooling setpoint}
        \label{fig:cooling_decision_tree}
    \end{subfigure}
    \vskip\baselineskip
    \begin{subfigure}{0.42\columnwidth}
        \includegraphics[width=\linewidth]{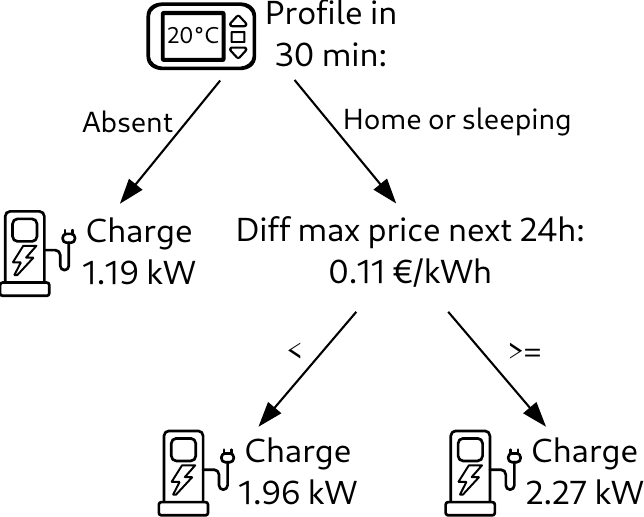}
        \caption{EV charger setpoint}
        \label{fig:charger_decision_tree}
    \end{subfigure}
    \hfill
    \begin{subfigure}{0.57\columnwidth}
        \includegraphics[width=\linewidth]{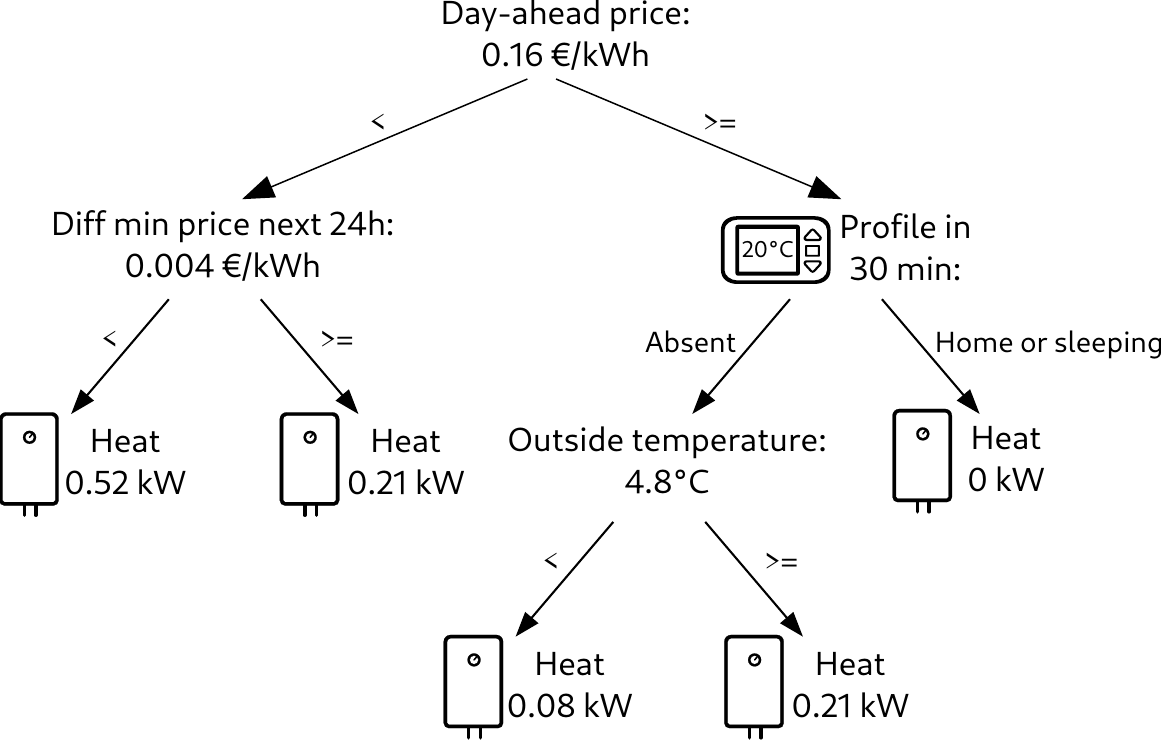}
        \caption{Water heater setpoint}
        \label{fig:water_heater_decision_tree}
    \end{subfigure}
    \caption{Decision trees for each controllable asset in the TreeC EMS for dwelling 1.}
    \label{fig:complete_dwelling}
    \Description{Four decision trees showing the control strategies for each controllable asset in dwelling 1. Each tree has nodes representing conditions based on features such as outdoor temperature, day-ahead price, occupancy profile, and tank temperature. The leaves of the trees indicate the setpoints for heating, cooling, EV charger power, and water heater power.}
\end{figure}

As an example, the TreeC EMS obtained after training for dwelling 1 is presented  in \cref{fig:complete_dwelling} as it is the first dwelling with all possible assets present (PV, heat pump, water heater and EV charger).

\begin{figure*}[ht]
    \centering
    \includegraphics[width=1.0\linewidth]{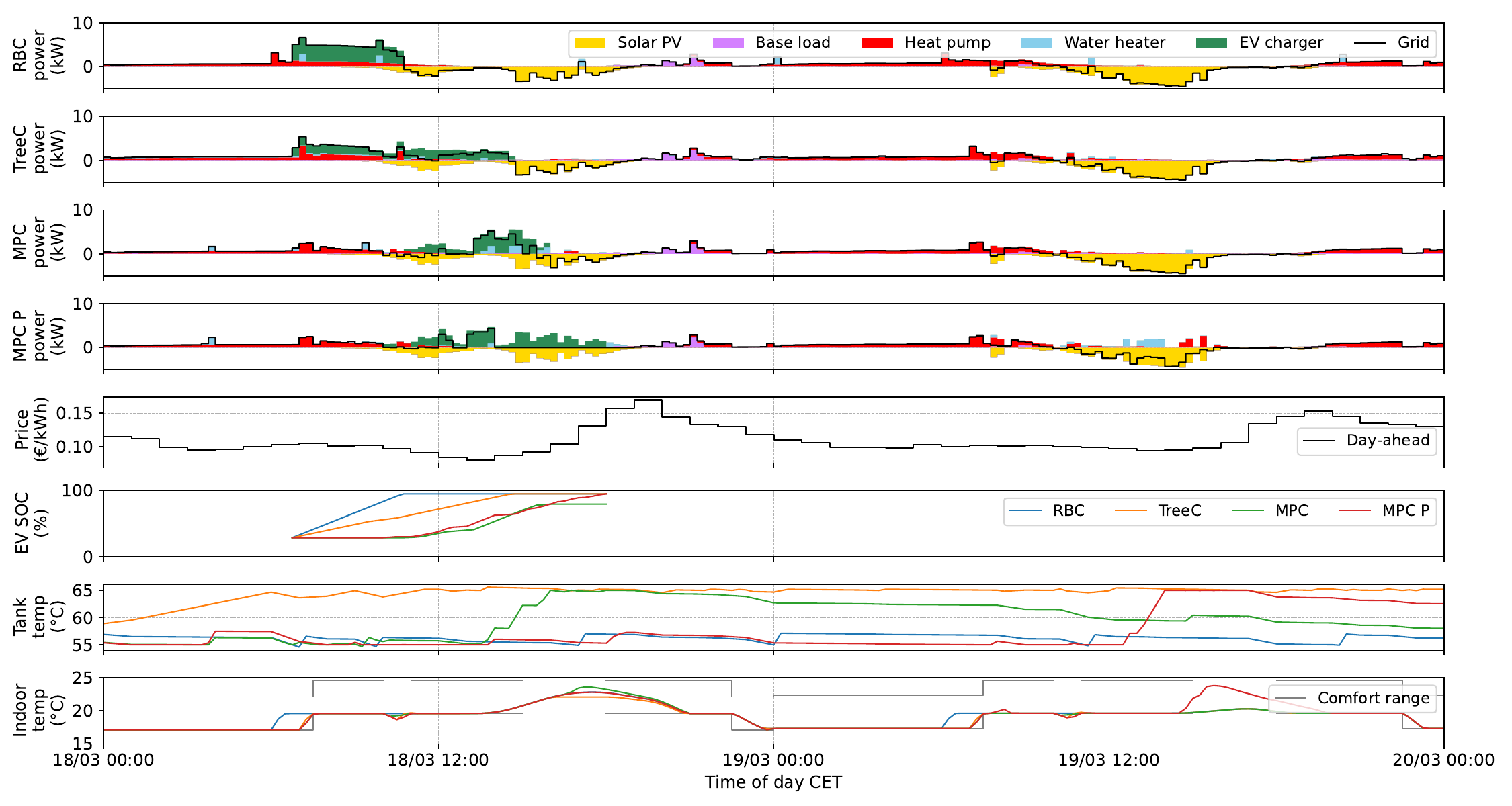}
    \caption{Illustration of the operation of the different EMSs over the weekend of 18-19 March 2023 for dwelling 1.
    The top four plots show the stacked power consumption and production of each asset, with the total grid connection power indicated in black.
    The four lower plots show the day-ahead electricity prices, the state of charge of the connected EV, the water heater's tank temperature, and the indoor temperature for each EMS.
    }
    
    \label{fig:example_weekend}
    \Description{Eight plots showing the operation of different EMSs over a weekend for dwelling 1. The top four plots display the power consumption and production of each asset, with the total grid connection power indicated in black. The four lower plots illustrate the day-ahead electricity prices, the state of charge of the connected EV, the water heater's tank temperature, and the indoor temperature for each EMS. The RBC EMS charges the EV at maximum power, while TreeC charges at a lower power to avoid peaks but still ensures full charge. MPC fails to fully charge the EV due to incorrect forecasting, while MPC P effectively utilizes excess PV production and low prices for charging. The water heater and heating strategies vary among EMSs based on electricity prices and PV production.}
\end{figure*}

The decision tree is interpretable for each of the set points it controls.
For the heating setpoint in \cref{fig:complete_dwelling}a, when the outdoor temperature is cold (below 4.8 °C), the EMS is precautious and matches the comfort range for the next 12 hours to not incur discomfort, otherwise it matches it for the next 30 minutes as this is more energy efficient. For the cooling setpoint in \cref{fig:complete_dwelling}b, when the day-ahead price is considered low enough (below 0.16\,€/kWh), the EMS matches the comfort range 7h30 minutes ahead, otherwise it just avoids that the indoor temperature increases too fast. For the EV charger in \cref{fig:complete_dwelling}c, the EMS charges between 1.96 kW and 2.27 kW when the occupancy profile is marked as home or sleeping in 30 minutes and at a lower power of 1.19 kW then occupancy profile is absent. The EV charger never charges at full power to avoid peaks. For the water heater in \cref{fig:complete_dwelling}d, the EMS heats the water more when the day-ahead price is considered low enough and doesn't heat at all when the occupancy profile is marked as home or sleeping in 30 minutes.

The figures and explanation highlight the interpretability of the TreeC EMS and also illustrate how it can coordinate the operation of different assets. For example, the EV charger tends to charge more when the water heater is not heating, helping to avoid simultaneous power peaks. This is achieved by both controllers using the same occupancy profile feature.

\section{Results}

\Cref{fig:example_weekend} shows how the different EMSs operate during the last weekend of winter for dwelling 1. This dwelling is a detached house in Flanders with all assets present: heat pump, water heater, EV charger, and PV system. The water heater has a 171 litre tank, the house is partially insulated (roof only), and the PV system has a capacity of 5 kWp.
This dwelling was chosen because it is the first with all assets present, and the selected weekend highlights an unsuccessful EV charging session under the MPC EMS. 

For the EV charging, RBC charges to the maximum power, TreeC charges at a lower power to avoid peaks but still high enough to ensure the EV is fully charged and MPC P charges the EV using all excess PV production and cheap day ahead prices. MPC however didn't fully charge the EV before departure. This is because it incorrectly forecasted that the EV would remain connected until 12:30 on the 19\textsuperscript{th} of March, whereas in reality, the EV departed at 18:00 on the 18\textsuperscript{th} of March. Because the day-ahead prices, and therefore injection prices, before 18:00 on the 18\textsuperscript{th} of March were higher than those before 12:30 on the 19\textsuperscript{th} of March, MPC planned to inject excess PV generation on the 18\textsuperscript{th} March and use the excess PV on 19 March to charge the EV. 
Mathematically this all makes sense, but since the prediction was incorrect, the EV was not fully charged by the time of departure contrary to the other EMSs.

For the domestic hot water, TreeC heats the water tank over almost the entire weekend because the price of electricity is mostly below 0.16\,€/kWh (see \cref{fig:complete_dwelling}d), MPC and MPC P make more use of the excess PV production to heat the water tank, and RBC keeps the water tank around the lower 55 °C bound.

For heating, all EMSs generally maintain the indoor temperature near the lower bound of the comfort range except for moments where the sun and outdoor temperatures heat the house naturally. MPC and MPC P take advantage of excess PV production to provide additional heating at certain times. RBC starts heating earlier in the morning than the other EMSs, which results in slightly lower energy efficiency.

\begin{figure*}[ht]
    \centering
    \begin{subfigure}{0.495\textwidth}
        \includegraphics[width=\textwidth]{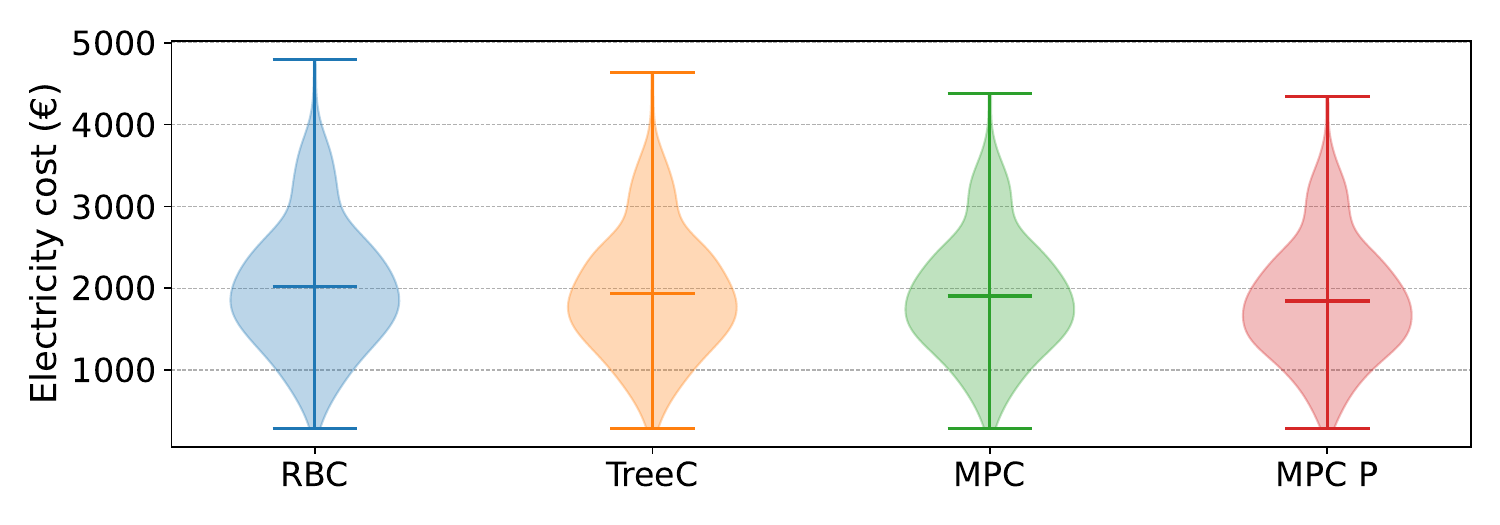}
        \caption{Yearly cost}
        \label{fig:total_opex}
    \end{subfigure}
    \hfill
    \begin{subfigure}{0.495\textwidth}
        \includegraphics[width=\textwidth]{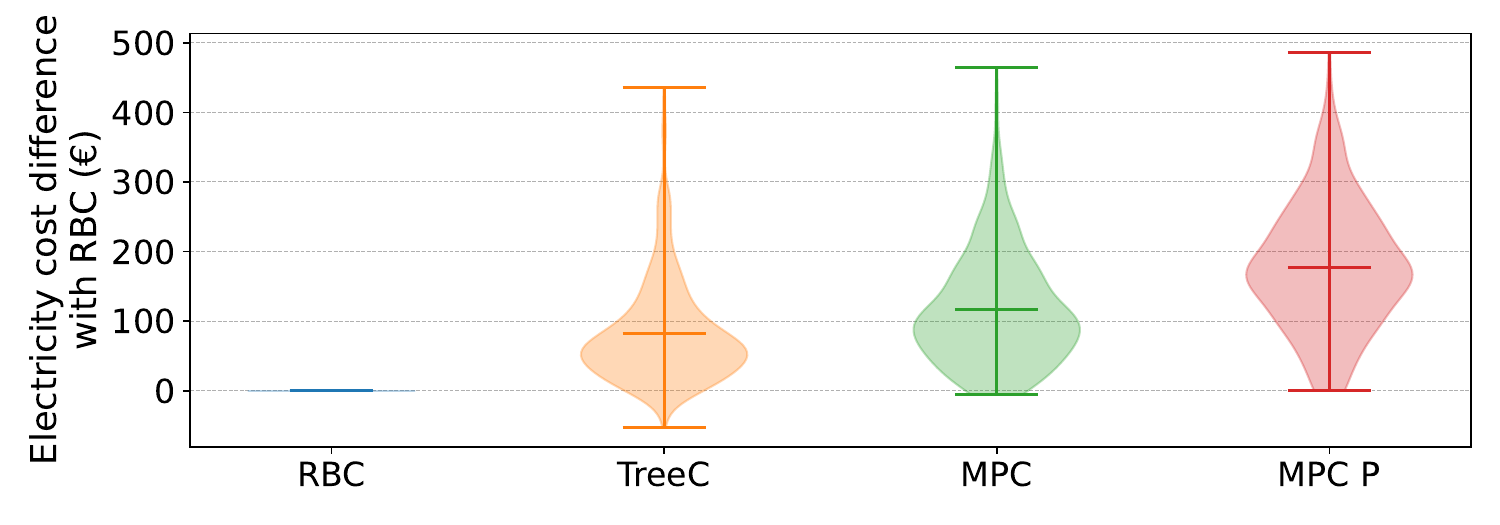}
        \caption{Difference with RBC}
        \label{fig:total_opex_diff}
    \end{subfigure}
    \begin{subfigure}{0.495\textwidth}
        \includegraphics[width=\textwidth]{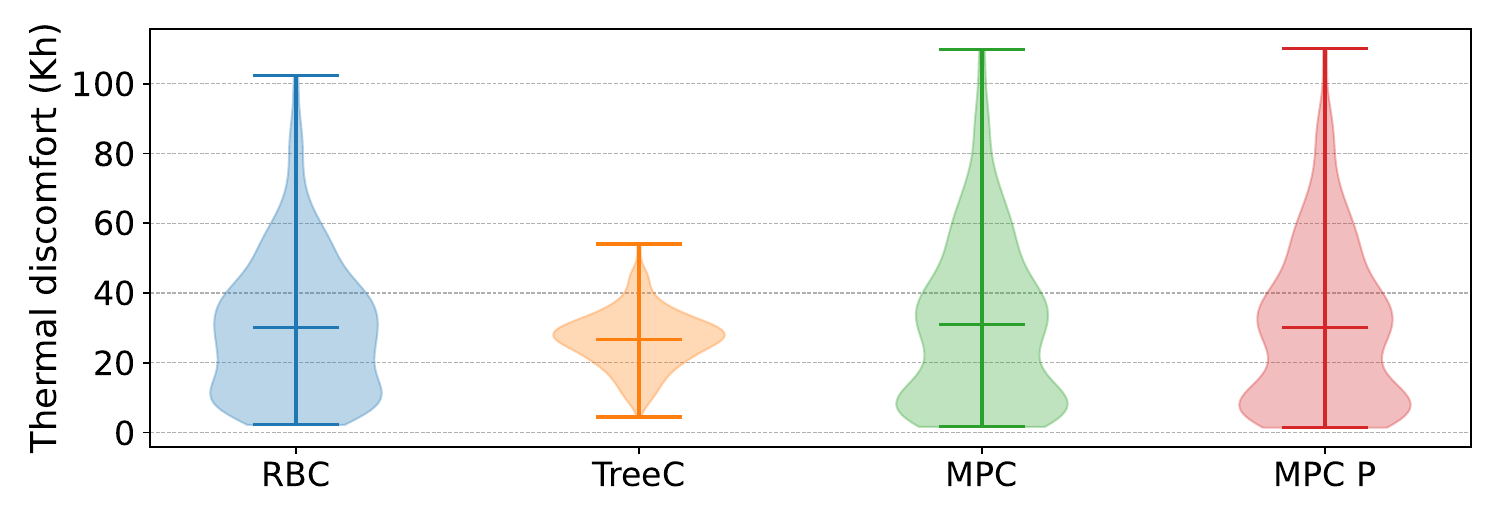}
        \caption{Thermal discomfort}
        \label{fig:discomfort_level}
    \end{subfigure}
    \hfill
    \begin{subfigure}{0.495\textwidth}
        \includegraphics[width=\textwidth]{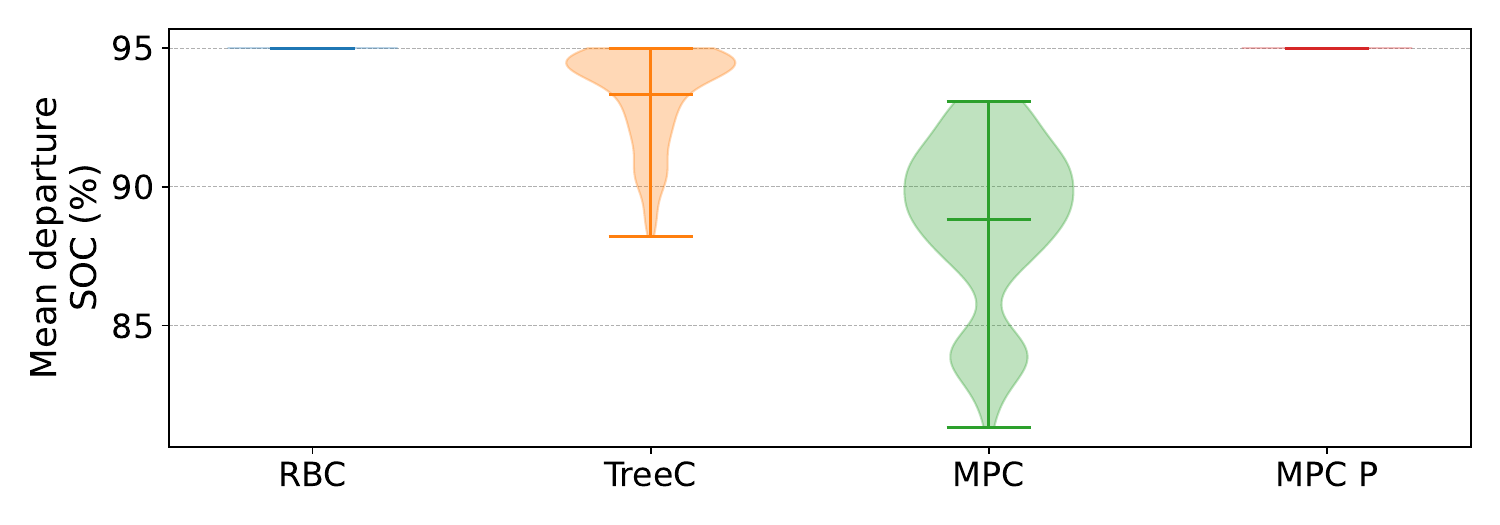}
        \caption{EV charge completion}
        \label{fig:avg_soc_final}
    \end{subfigure}
    
    \caption{Comparison of key performance metrics for the different EMSs: (a) total yearly cost, (b) difference in electricity cost with the RBC EMS, (c) thermal discomfort, and (d) mean departure state of charge for EV charging sessions.
    For each EMS, violin plots (a) and (b) include 500 dwellings, (c) includes the 447 dwellings with a heat pump, and (d) includes the 344 dwellings with an EV charger. Each violin plot shows the distribution of results across all dwellings. The width of the violin at each value reflects the relative frequency of dwellings with that result. The horizontal dashes within each violin indicate the minimum, mean, and maximum values for each EMS.
    }
    \label{fig:performance_comparison}
    \Description{Four violin plots comparing key performance metrics for different EMSs across multiple dwellings. The first plot shows the total yearly cost, indicating a hierarchy in performance with RBC having the highest costs and MPC P the lowest. The second plot illustrates the difference in electricity cost compared to RBC, showing significant cost reductions for TreeC, MPC, and MPC P. The third plot presents thermal discomfort levels, with all EMSs achieving similar mean discomfort levels but TreeC maintaining lower maximum discomfort. The fourth plot displays the mean departure state of charge for EV charging sessions, highlighting that TreeC outperforms MPC in ensuring higher charge levels at departure. Each plot includes horizontal dashes indicating minimum, mean, and maximum values for each EMS.}
\end{figure*}

\begin{figure*}[ht]
    \centering
    \begin{subfigure}{0.495\textwidth}
        \includegraphics[width=\textwidth]{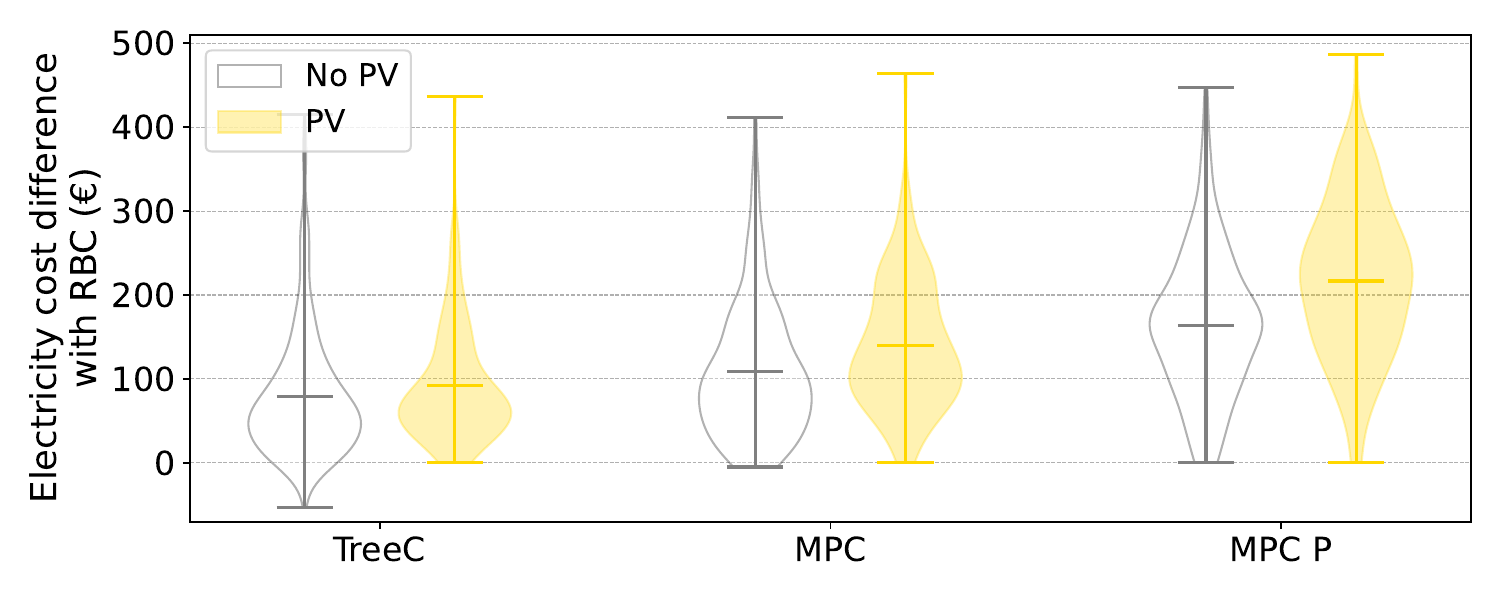}
        \caption{PV}
        \label{fig:pv_results}
    \end{subfigure}
    \hfill
    \begin{subfigure}{0.495\textwidth}
        \includegraphics[width=\textwidth]{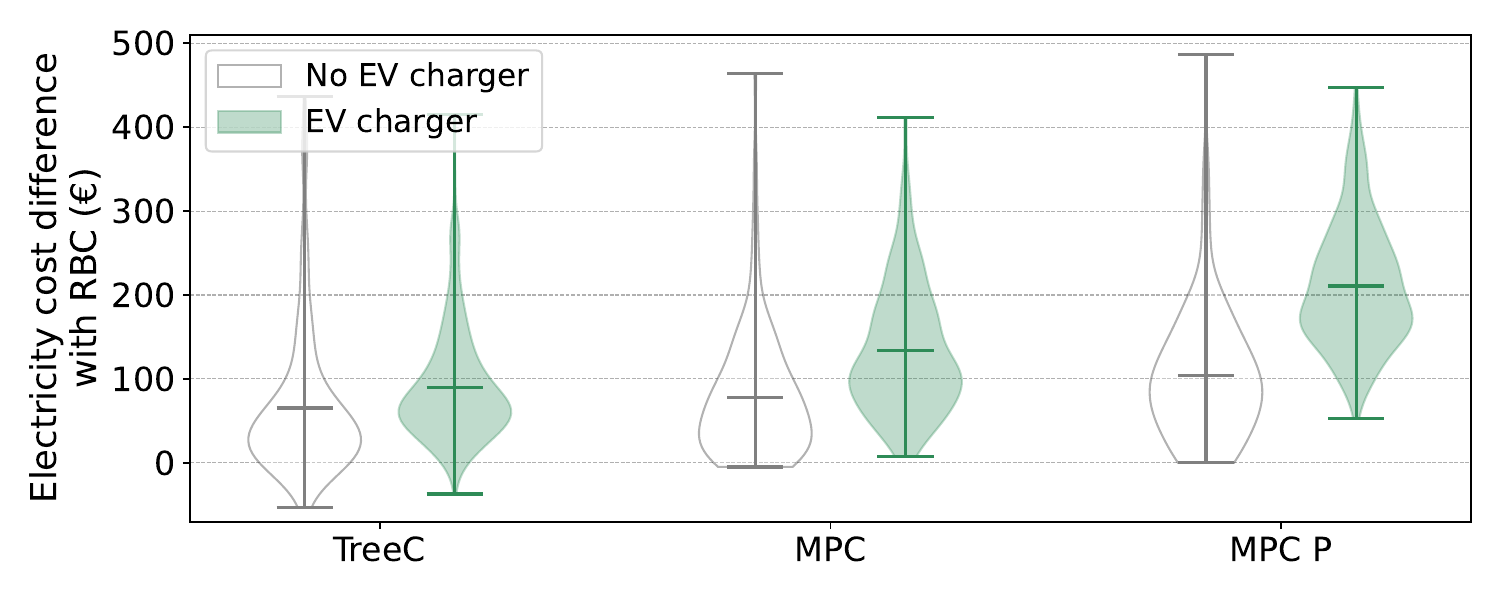}
        \caption{EV charger}
        \label{fig:home_charger_results}
    \end{subfigure}
    \begin{subfigure}{0.495\textwidth}
        \includegraphics[width=\textwidth]{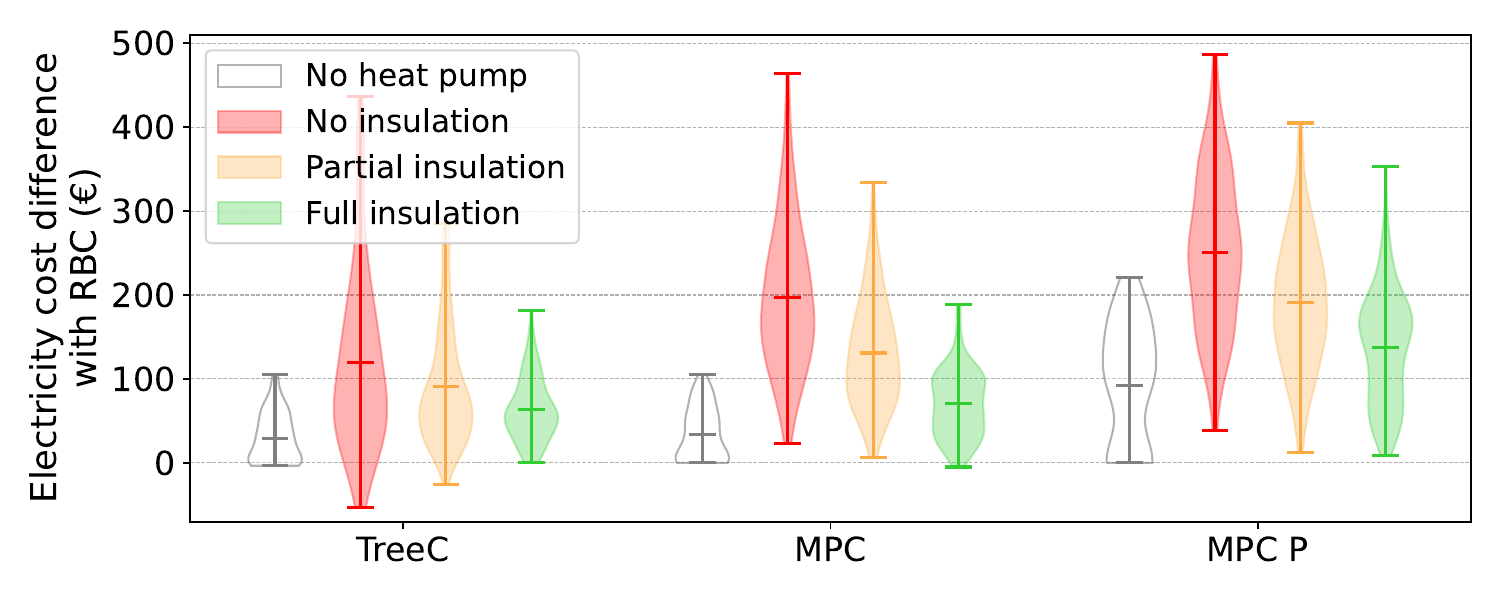}
        \caption{Insulation}
        \label{fig:insulation_results}
    \end{subfigure}
    \hfill
    \begin{subfigure}{0.495\textwidth}
        \includegraphics[width=\textwidth]{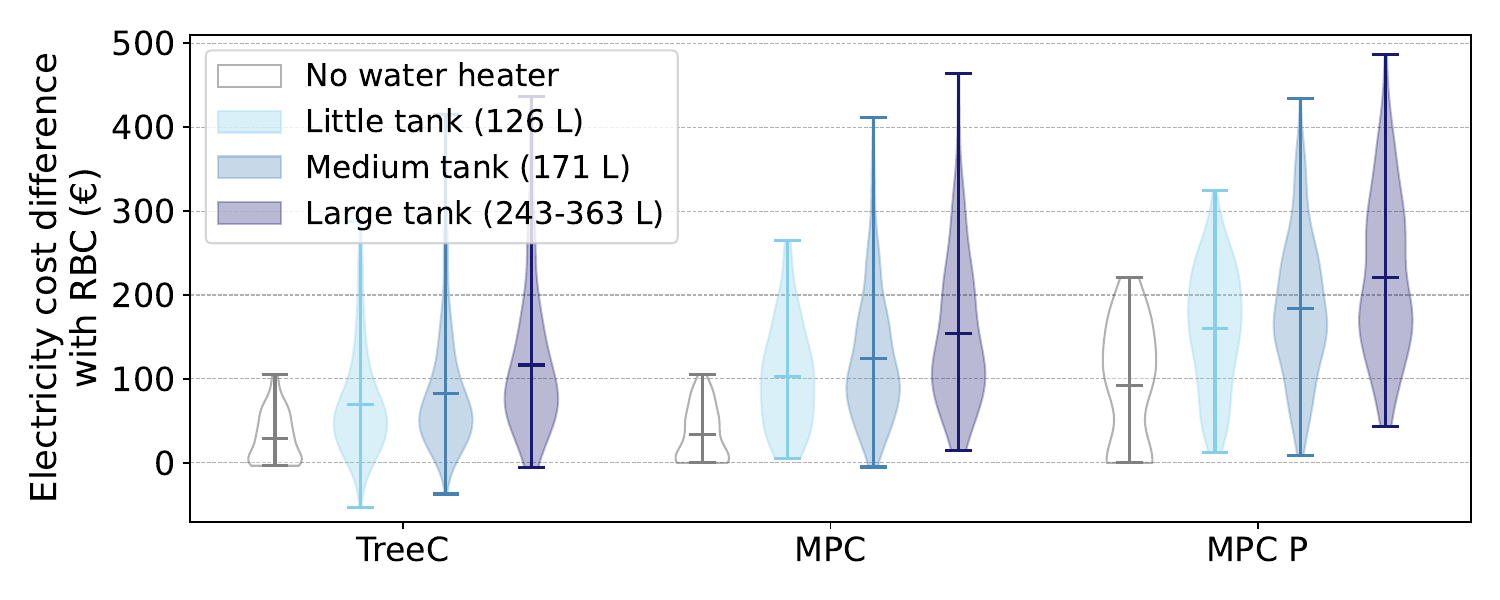}
        \caption{Tank volume}
        \label{fig:water_heater_volume_results}
    \end{subfigure}

        \caption{Comparison of EMS performance for different dwelling features: (a) PV presence, (b) EV charger presence, (c) insulation level, and (d) water heater's tank volume %
    }
    \label{fig:dwelling_features_results}
    \Description{Five subplots comparing EMS performance based on different dwelling features. Each subplot presents violin plots showing the distribution of cost savings relative to the RBC EMS for each feature category. The features analyzed include the presence of PV systems, EV chargers, insulation levels, water heater tank volumes, and the number of assets in the dwelling. Each violin plot illustrates the variability in EMS performance across dwellings with the specified feature, with horizontal dashes indicating minimum, mean, and maximum values for each EMS.}
\end{figure*}

Results in \cref{fig:performance_comparison} show different metrics to compare EMSs across the 500 dwellings.
\Cref{fig:performance_comparison}a shows the total yearly cost for each dwelling. The results indicate a hierarchy in performance: the RBC EMS has the highest costs, followed by TreeC, then MPC, with MPC P achieving the lowest costs overall. There is considerable variation in total costs across dwellings, reflecting the impact of factors such as insulation quality, and the presence of EV chargers or PV systems on overall energy consumption.

To provide a clearer comparison of the EMS performance relative to RBC, \cref{fig:performance_comparison}b presents the reduction in total yearly cost for each EMS compared to the RBC EMS. The results show that TreeC achieves a mean cost reduction of 82.5\,€, MPC achieves 116.6\,€, and the upper bound MPC P achieves 177.4\,€.
TreeC and MPC rarely achieve a worse cost than RBC, which is likely counteracted by an increase in thermal comfort.

Figure~\ref{fig:discomfort_level} shows the thermal discomfort for each EMS with all EMSs achieving similar mean discomfort levels between 26 and 31 Kh. TreeC manages to keep all dwellings below 55 Kh discomfort while the other EMSs have some dwellings with discomforts up to 111 Kh.
Overall even 111 Kh can be considered acceptable especially since some dwellings are not insulated at all and this only represents 111 hours in the year when the indoor temperature is 1 °C above or below the comfort range with the rest of the year being within the comfort range.

Figure~\ref{fig:avg_soc_final} shows the mean departure SOC for the EV charging sessions across all dwellings with an EV charger. The original dataset assumed the departure charging SOC to be 95\% for all sessions so it is normal that RBC and MPC P always achieve this value.
For this particular metric TreeC performs better than MPC achieving a mean of 93.3\% departure SOC compared to 88.8\% for MPC.
This shows that TreeC is able to charge the EV more effectively than MPC but the average departure SOC does not always give a complete representation of the charging performance.

\Cref{fig:dwelling_features_results} shows the impact of different dwelling features on the EMS performance. Most results are as expected: \cref{fig:dwelling_features_results}a shows that an EMS makes better improvements when the dwelling has a PV system, \cref{fig:dwelling_features_results}b shows the same for EV chargers and figs.~\ref{fig:insulation_results} and \ref{fig:water_heater_volume_results} shows the same for central heating. %

Interestingly \cref{fig:dwelling_features_results}c shows that more insulation has negative impact on the EMSs improvement compared to RBC. This is certainly because dwellings with more insulation have a lower energy consumption and therefore the EMSs have less opportunity to reduce costs.
For the water tank size, \cref{fig:dwelling_features_results}d shows a positive correlation between the tank size and the EMS performance.

\begin{figure}[ht]
    \centering
    \includegraphics[width=\columnwidth]{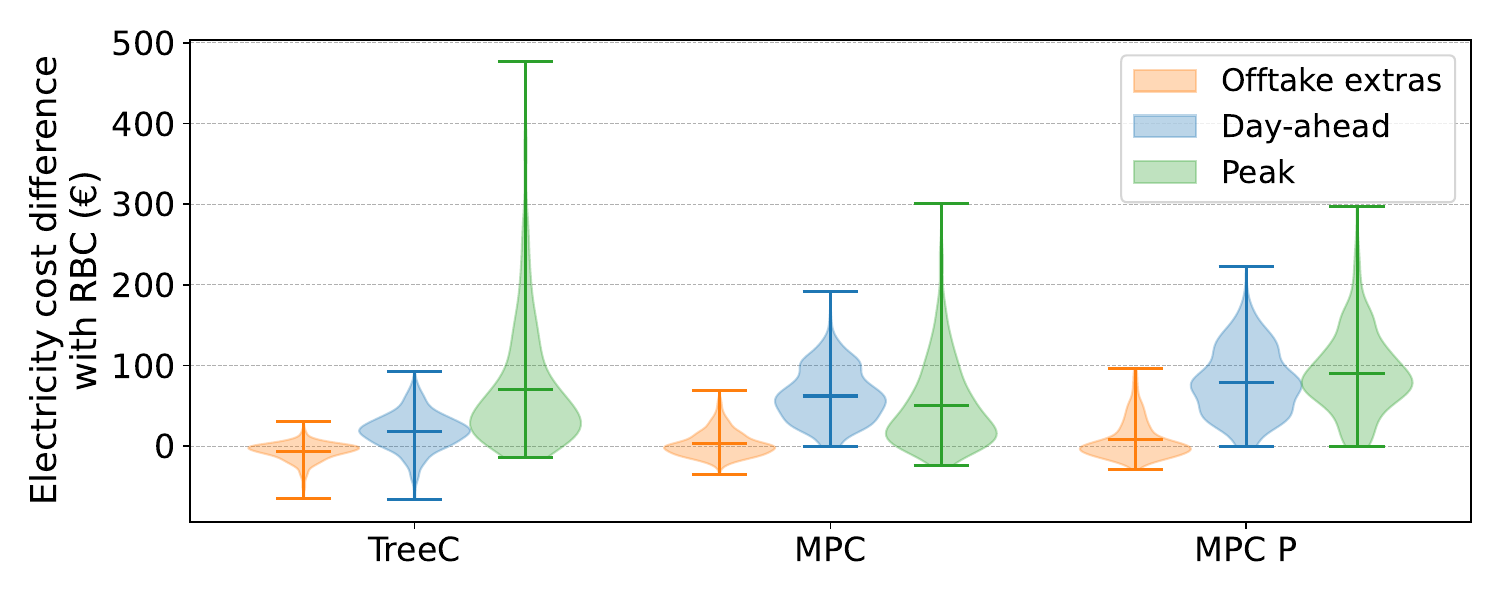}
    \caption{Breakdown of cost savings relative to the RBC EMS for each cost component defined in \cref{eq:tot_cost_usefull}. Offtake extras costs improvement reflects better energy efficiency, day-ahead costs improvement reflects better hourly price optimization, and peak costs improvement reflects reductions in peak grid power. Each box plot summarizes the results across the 500 dwellings.}
    \label{fig:opex_diff_separate}
    \Description{Box plots breaking down cost savings relative to the RBC EMS for each cost component: offtake extras costs, day-ahead costs, and peak costs. Each box plot summarizes the distribution of cost savings across 500 dwellings for each EMS. The offtake extras costs improvement shows minimal savings, indicating limited gains in energy efficiency. The day-ahead costs improvement highlights significant savings achieved by MPC and MPC P, reflecting their superior hourly price optimization capabilities. The peak costs improvement indicates that TreeC slightly outperforms MPC in reducing peak grid power consumption. Each box plot includes horizontal lines representing the median, interquartile range, and potential outliers for each EMS.}
\end{figure}

To understand better how the EMSs achieve cost reductions, \cref{fig:opex_diff_separate} breaks down the savings by cost component. 

Only minimal savings are achieved in the offtake extras costs related to energy efficiency. This is expected, as the RBC is already quite energy efficient for water and space heating. Efficiency gains can be realized by directly consuming electricity from PV, but 74.2\% of dwellings do not have a PV system.

MPC and MPC P achieve significant cost saving in the day-ahead costs, this is the main cost component that gives these methods their edge over TreeC. This can be explained by MPC based techniques having perfect hourly knowledge of the future day-ahead prices and being able to schedule assets based on these prices more effectively than TreeC which relies on incomplete knowledge of the future prices and a less sophisticated decision tree control model.

For the peak cost component, TreeC slightly outperforms MPC. During training, TreeC learns to minimize peak costs by coordinating asset operation and selecting lower power setpoints, thereby avoiding large simultaneous peaks. While MPC also attempts to coordinate assets to reduce peaks, prediction errors can occur due to forecast or model inaccuracies and result in unexpected peaks, which may explain its lower performance compared to TreeC in this area.

\begin{figure}[ht]
    \centering
    \includegraphics[width=1.0\columnwidth]{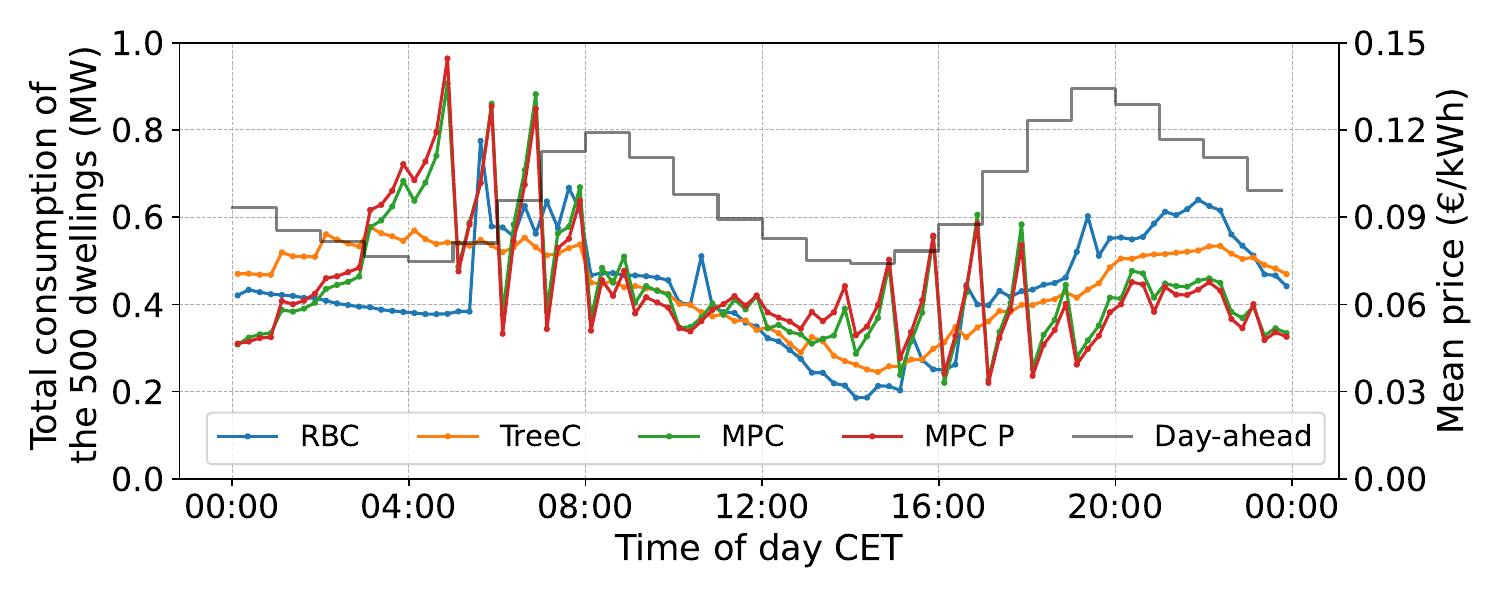}
    \caption{Yearly mean of the total grid power consumption of the 500 dwellings for each time step of the day. Each EMS is represented as well as the mean day-ahead electricity prices ($Vd_t$) for the year 2023.
    }
    \label{fig:average_daily_grid_power}
    \Description{Line plot showing the yearly mean of total grid power consumption for 500 dwellings across different EMSs for each time step of the day. The plot includes lines for each EMS (RBC, TreeC, MPC, MPC P) and a line representing the mean day-ahead electricity prices for 2023. The x-axis represents the time of day in hours, while the y-axis indicates the total grid power consumption in kW. The plot illustrates how each EMS manages grid power consumption throughout the day in relation to electricity pricing.}
\end{figure}

Another relevant aspect to consider is how the peak and hourly day-ahead costs influence the total grid power consumption of the 500 dwellings. \Cref{fig:average_daily_grid_power} shows on 24 hours the mean total grid power for each EMS. The MPC and MPC P EMSs show almost identical average behaviour. Overall TreeC and the MPC EMSs consume more than RBC during hours with lower day ahead prices, and all manage reduce the evening RBC total peak. 
However, the MPC EMSs show a clear peak for the last time step of the hours between 04:00 and 08:00 and between 14:00 and 18:00 which are the times the dwelling heat up to reach a higher comfort temperature. This is because the MPC EMSs calculates that it is better to heat the dwelling at the last time step of the hour to gain advantage of the low day-ahead price of that hour and at the same time loose as little as possible of the heat bought at that price. Buying at an earlier time step of that hour would result in more of the bought heat being lost to the exterior. 
The TreeC EMSs didn't learn this behaviour and therefore has a much smoother total grid power consumption profile. 

The peaks caused by the MPC EMSs are problematic for the grid because: 1) they can cause local congestion, requiring costly grid reinforcement, and 2) the sharp, short-lived spikes in aggregate demand increase intra-hour variability, making grid balancing more challenging for operators. This undermines the intent of hourly pricing signals, which are designed to encourage demand shifting across entire hours rather than concentrating consumption at specific moments within those hours.

\section{Discussion}

When analysing all the results, TreeC and MPC have their strengths and weaknesses.
TreeC is more interpretable (see \cref{fig:complete_dwelling}), achieves higher EV charge completion rates, and results in a smoother total grid power consumption profile when all dwellings are equipped with the same EMS. However, MPC obtains more cost reductions, which is arguably the most important metric for evaluating EMS performance.

To improve the cost reduction of TreeC, future work should focus on improving its ability to reduce the day-ahead costs. This can be done by retraining every month the EMS to adapt to the changing day-ahead prices, adding more price related inputs that can be selected by the decision trees, or using more sophisticated action on the leaf nodes of the decision trees. Improving the training time of TreeC through reduced simulation speed, improved training methodology, or transfer learning is also necessary to make it more practical for real-world applications.

\begin{table}[H]
    \centering
    \caption{Mean absolute error (MAE) of two forecasts for the EV charging sessions. Values show the mean and standard deviation of MAE across all dwellings with an EV charger for 2023.}
    \small
    \begin{tabular}{|l|l|c|}
        \hline
        Forecast method & Predicted value & MAE (Mean  $\pm$ Std. dev.) \\
        \hline
        K-NN regression & Connection time & 7h14min $\pm$ 4h33min \\
        Repeat last & Connection time & 9h14min $\pm$ 5h7min \\
        K-NN regression & SOC demand  & 12.8\% $\pm$ 3.4\% \\ 
        Repeat last & SOC demand  & 16.4\% $\pm$ 5.4\% \\
        \hline
    \end{tabular}
    \label{tab:ev_charging_stats}
\end{table}

The results for MPC P indicate that improved forecasting can enhance both cost savings and EV charge completion. However, accurately predicting EV charging sessions remains challenging due to the high variability observed in table~\ref{tab:ev_charging_profile_stats} and the substantial mean absolute errors (MAE) for both k-NN regression and "repeat last" forecasts in \cref{tab:ev_charging_stats} exceeding 7 hours for connection time and 12\% for SOC demand. To address this, future work should consider multi-scenario or stochastic MPC approaches that explicitly account for the uncertainty in EV charging behaviour, as conventional forecasting methods struggle with this aspect.

The mean annual savings compared to the RBC EMS are 82.5\,€ for TreeC, 116.6\,€ for MPC, and 177.4\,€ for MPC P. These savings are relatively modest compared to the cost of buying, installing, maintaining, and subscription fees of current home EMS solutions on the market (between 280\,€ and 680\,€ to buy and 0\,€ and 360\,€ per year for subscription fees \cite{maakjemeterslimAlleEMSProducten2025}). Specific cases can justify an EMS implementation as the results do show some cost reduction going up to 464\,€ are possible, but these are not the norm.

The results of MPC P also suggest that algorithmic improvements alone are unlikely to yield significantly greater benefits than the 177.4\,€ mean cost reduction. 

The code and data for this study are open-source, enabling other researchers to benchmark their EMS solutions in the same context. All materials are available at:
\url{https://github.com/EVERGi/should_belgians_install_advanced_ems_paper}.

In addition to algorithmic advances, several external factors could improve the commercial attractiveness of advanced EMSs:

\begin{itemize}
\item More volatility in electricity prices. Having more extreme low and high prices would increase the potential savings of shifting consumption towards low price periods.
\item Allowing participation in balancing markets. This would create additional revenue streams for households providing flexibility to the grid.
\item A wider adoption of home batteries. The additional flexibility provided by batteries could increase the potential savings from advanced EMSs.
\item Wider adoption of standardized protocols in large electrical appliances. Standardization would lower EMS integration costs by making it easier to connect and control devices, potentially allowing simple plug-and-play EMS installation.
\end{itemize}

\section{Conclusion}

This study presented a comparison of three EMSs for electrified residential buildings in Belgium: a rule-based controller, a model predictive control EMS, and a decision tree-based EMS called TreeC. The study evaluated the performance across one year simulations of 500 dwellings with characteristics representative of current Belgian dwellings but where the fossil fuel based assets for heating, hot water and car transportation are replaced by electric equivalents. 

On average, TreeC and MPC achieved cost reductions of 82.5\,€ and 116.6\,€, respectively, compared to the RBC EMS, which had a mean total cost of 2021\,€.
MPC obtained better cost reduction but worse EV charge completion with an average of 88.8\% SOC at departure compared to 93.3\% for TreeC with a full charge set at 95\% SOC.
The total power consumption of the 500 dwellings together obtained large volatility from one time step to the next when all dwellings use MPC, while TreeC obtained a smoother total power consumption profile.

The main recommended improvement for TreeC is to enhance its ability to reduce hourly day-ahead costs. This could be achieved by retraining the EMS more frequently to adapt to changing electricity over time or by making the decision tree model more responsive to hourly price signals. For MPC, future work should focus on addressing the uncertainty in EV charging sessions, for example by adopting stochastic or multi-scenario MPC approaches.

The results show that TreeC and MPC performed better than RBC in terms of cost for dwellings with PV systems, with EV chargers, and with larger water tanks,
which is expected, it also worked better for dwellings with less insulation which is a less intuitive result.
Nonetheless, compared to the default rule-based operation of assets (RBC), advanced EMSs offer modest cost savings on average (82.5\,€ and 116.6\,€) that generally do not justify the investment (280--680\,€) and maintenance costs (0--360\,€ subscription costs per year) of the current available commercial options \cite{maakjemeterslimAlleEMSProducten2025}.
Further algorithmic improvements are unlikely to deliver significantly greater benefits than a mean cost reduction of 177.4\,€ achieved by the MPC with perfect forecast (MPC P) EMS.
Increasing the potential cost savings, and thus the attractiveness of advanced EMSs, could be achieved by modifying electricity pricing structures, enabling participation in flexibility markets, reducing EMS investment and maintenance costs, or integrating home batteries.

\begin{acks}
This work has been supported by the ECOFLEX project funded by the FPS economy, S.M.E.s, Self-employed and Energy, Belgium and by the Flemish Government and Flanders Innovation \& Entrepreneurship (VLAIO) through the Flux50 project LEVEL \linebreak(HBC.2025.0421).
\end{acks}

\bibliographystyle{ACM-Reference-Format}
\bibliography{bibfile}

\clearpage

\appendix

\makeatletter
\@addtoreset{figure}{section}
\@addtoreset{table}{section}
\@addtoreset{equation}{section}
\makeatother
\renewcommand{\thefigure}{\thesection.\arabic{figure}}
\renewcommand{\thetable}{\thesection.\arabic{table}}
\renewcommand{\theequation}{\thesection.\arabic{equation}}

\section{Justification for representative set of Belgian dwellings}
\label{sec:representative_dwellings}

\subsection{Dwelling characteristics}
\label{sec:dwelling_characteristics}

Different statistics were collected to build the set of 500 representative Belgian dwellings.
This section goes through each of them giving the source and explanation when necessary. In the following tables, each cell shows the percentage of all dwellings that have both the characteristic specified by the row and the characteristic specified by the column.

To assign characteristics to the 500 representative dwellings, the number of dwellings with each characteristic is calculated by multiplying the relevant percentage by 500. For example, according to \cref{tab:stats_region}, 10.6\% of dwellings are in Brussels, 57.4\% in Flanders, and 32.0\% in Wallonia, resulting in 53 dwellings in Brussels, 287 in Flanders, and 160 in Wallonia. This establishes the initial pool of unassigned characteristics.

Assignment proceeds sequentially, starting with the most independent characteristic (region), and moving towards the most dependent (number of bedrooms and surface area) as illustrated by \cref{fig:dwelling_flow_chart}. For each dwelling, characteristics are sampled randomly from the remaining pool, using any already assigned characteristics as conditional constraints. This process continues until all dwellings have a complete set of characteristics, and the overall distribution matches the target statistics.

The percentage for the number of dwellings per region (\cref{tab:stats_region}), the presence of central heating (\cref{tab:stats_heating}) and the number of rooms (\cref{tab:stats_rooms}) are all obtained from the 2021 Belgian Census which gives the total number of dwellings fitting each characteristic without need for extrapolation \cite{statbelCensus2021Nombre2021}.

The number of bedrooms data from \cref{tab:stats_bedrooms} comes from an extrapolation of \citet{meirelesDomesticHotWater2022}. 
The source gives the number of bedrooms of 9\,206 Belgian apartments. 
To fit better to the global Belgian situation with 66\% houses (see \cref{tab:stats_construction_type}), the total proportion of 0 and 1 bedroom dwellings was reduced by 25\% and the total proportion of 3 to 6 bedroom dwellings was increased by 47.3\% to match the reduction of the 0 and 1 bedroom dwellings.
The correlation between the number of bedrooms and the number of rooms is established using common sense with the total bedroom and total room  proportions  as a constraint (see \cref{tab:stats_rooms}).
\citet{meirelesDomesticHotWater2022} indicates that the number of bedrooms in a dwelling is a good indicator of average water consumption. 
The number of bedrooms is used solely to determine the appropriate hot water consumption profile to assign to the dwelling, as detailed in \cref{sec:hot_water}.

The surface area data from \cref{tab:stats_surface_area} comes from tables IV.24 and IV.28 of \citet{vannesteLogementBelgique2007}.
Table IV.24 gives the number of dwellings with a certain surface area per region. 
Table IV.28 gives the number of dwellings with a certain number of rooms per region.
The correlation between the number of rooms and the surface area is determined to ensure that the final proportions per region closely match those presented in tables IV.24 and IV.28 of \citet{vannesteLogementBelgique2007}.

The garage and car ownership data from \cref{tab:stats_home_charger} comes from page 68 of \citet{hubertMobiliteQuotidienneBelges2003}.
The source gives the proportion of dwellings that have a garage in each region and the proportion of people with a garage that have at least one car.
\Cref{tab:stats_home_charger} shows the combination of these two proportions, it is used to determine the proportion of dwellings that would have a EV charger in each region.

The PV data from \cref{tab:stats_pv} comes from different sources for the different regions.
The number of photovoltaic installations in Brussels and Flanders comes from \citet{brugelEnergiesRenouvelablesRegion2024} and \citet{vlaamsenergie-enklimaatagentschapZonnepanelenVlaanderen2024} respectively. For Wallonia, the proportion of dwellings with a PV installation is given by \citet{lallemand16MenagesWallons2023}.

The construction type data in \cref{tab:stats_construction_type} combines two sources: \citet{statbelCensus2021Nombre2021b}, which provides the proportion of houses and apartments in each region, and Figure IV.13 from \citet{vannesteLogementBelgique2007}, which details the breakdown of terraced, semi-detached, and detached constructions by region.

The insulation data in \cref{tab:stats_insulation} is based on \citet{recticelinsulationBarometreLisolation20202020}, which reports the proportions of dwellings with roof, wall, and ground insulation in each region. However, the source does not specify the exact overlap between these insulation levels (e.g., how many dwellings have both wall and roof insulation, or all three types). To construct the final table, the following assumptions were made: dwellings with ground insulation also have wall and roof insulation, and dwellings with wall insulation also have roof insulation. This approach ensures that the total proportions match the source data while providing a reasonable estimate of the distribution of combined insulation levels.

\begin{table}[H]
    \centering
    \caption{Distribution of dwellings per region \cite{statbelCensus2021Nombre2021}}
    \begin{tabular}{|c | c | c|}
    \hline
    Brussels & Flanders & Wallonia \\
    \hline
    10.6\% & 57.4\% & 32.0\% \\
    \hline
    \end{tabular}
    \label{tab:stats_region}
\end{table}

\begin{table}[H]
    \centering
    \caption{Distribution of dwellings with central heating \cite{statbelCensus2021Nombre2021}}
    \small
    \begin{tabular}{|c | c | c | c||c|}
    \hline
     & Brussels & Flanders & Wallonia & Total \\
    \hline
    With central heating & 10.1\% & 51.9\% & 27.4\% & 89.4\% \\
    \hline
    Without central heating & 0.5\% & 5.5\% & 4.6\% & 10.6\% \\
    \hline
    \end{tabular}
    \label{tab:stats_heating}
\end{table}

\begin{table}[H]
    \centering
    \caption{Distribution of dwellings per number of rooms \cite{statbelCensus2021Nombre2021a}}
    \begin{tabular}{|c | c | c | c||c|}
    \hline
     & Brussels & Flanders & Wallonia & Total \\
    \hline
    1 room & 0.3\% & 0.5\% & 0.4\% & 1.2\% \\
    \hline
    2 rooms & 0.8\% & 1.6\% & 1.0\% & 3.4\% \\
    \hline
    3 rooms & 2.1\% & 4.6\% & 2.5\% & 9.2\% \\
    \hline
    4 rooms & 3.2\% & 10.0\% & 5.1\% & 18.3\% \\
    \hline
    5 rooms & 1.8\% & 14.0\% & 7.4\% & 23.2\% \\
    \hline
    6 rooms & 0.9\% & 12.6\% & 7.1\% & 20.6\% \\
    \hline
    7 rooms & 0.5\% & 7.3\% & 4.1\% & 11.9\% \\
    \hline
    8 rooms & 0.4\% & 3.6\% & 2.3\% & 6.3\% \\
    \hline
    9+ rooms & 0.6\% & 3.2\% & 2.1\% & 5.9\% \\
    \hline
    \end{tabular}
    \label{tab:stats_rooms}
\end{table}

\begin{table}[H]
    \centering
    \caption{Distribution of dwellings per number of bedrooms \cite{meirelesDomesticHotWater2022}}
    \setlength{\tabcolsep}{4pt}
    \scriptsize
    \begin{tabular}{|c | c | c | c | c | c | c | c | c | c||c|}
    \hline
    \begin{tabular}[c]{@{}c@{}}Number of\\rooms\end{tabular} & 1 & 2 & 3 & 4 & 5 & 6 & 7 & 8 & 9+ & Total \\
    \hline
    0 bedroom & 0.4\% & 1.2\% & 2.5\% & 2.5\% & 2.0\% & 0.3\% & - & - & - & 8.9\% \\
    \hline
    1 bedroom & 0.8\% & 2.2\% & 5.6\% & 5.6\% & 4.6\% & 1.2\% & - & - & - & 20.0\% \\
    \hline
    2 bedrooms & - & - & 1.1\% & 8.2\% & 12.3\% & 10.3\% & 5.1\% & 2.1\% & 2.0\% & 41.1\% \\
    \hline
    3 bedrooms & - & - & - & 2.0\% & 4.0\% & 6.5\% & 5.1\% & 2.2\% & 2.1\% & 21.9\% \\
    \hline
    4 bedrooms & - & - & - & - & 0.3\% & 2.2\% & 1.4\% & 1.3\% & 1.0\% & 6.2\% \\
    \hline
    5 bedrooms & - & - & - & - & - & 0.1\% & 0.3\% & 0.6\% & 0.5\% & 1.5\% \\
    \hline
    6 bedrooms & - & - & - & - & - & - & - & 0.1\% & 0.3\% & 0.4\% \\
    \hline
    \end{tabular}
    \label{tab:stats_bedrooms}
\end{table}

\begin{table}[H]
    \centering
    \caption{Distribution of dwellings per surface area \cite{vannesteLogementBelgique2007}}
    \setlength{\tabcolsep}{4pt}
    \scriptsize
    \begin{tabular}{|c | c | c | c | c | c | c | c | c | c||c|}
    \hline
    \begin{tabular}[c]{@{}c@{}}Number of\\rooms\end{tabular} & 1 & 2 & 3 & 4 & 5 & 6 & 7 & 8 & 9+ & Total \\
    \hline
    <35m² & 1.1\% & 1.7\% & 0.9\% & - & - & - & - & - & - & 3.7\% \\
    \hline
    35-54m² & 0.1\% & 1.0\% & 2.8\% & 5.5\% & 4.6\% & 3.1\% & - & - & - & 17.1\% \\
    \hline
    55-84m² & - & 0.7\% & 3.7\% & 7.3\% & 7.0\% & 5.1\% & 1.2\% & - & - & 25.0\% \\
    \hline
    85-104m² & - & - & 1.8\% & 3.7\% & 7.0\% & 6.2\% & 3.6\% & 0.5\% & 0.5\% & 23.3\% \\
    \hline
    105-124m² & - & - & - & 1.8\% & 2.3\% & 4.1\% & 3.6\% & 1.6\% & 1.5\% & 14.9\% \\
    \hline
    >124m² & - & - & - & - & 2.3\% & 2.1\% & 3.5\% & 4.2\% & 3.9\% & 16.0\% \\
    \hline
    \end{tabular}
    \label{tab:stats_surface_area}
\end{table}
    
\begin{table}[H]
    \centering
    \caption{Distribution of dwellings with a garage and a car \cite{hubertMobiliteQuotidienneBelges2003}}
    \begin{tabular}{|c | c | c | c||c|}
    \hline
     & Brussels & Flanders & Wallonia & Total \\
    \hline
    With garage and car & 3.8\% & 42.7\% & 22.4\% & 68.9\% \\
    \hline
    Without garage or car & 6.8\% & 14.7\% & 9.6\% & 31.1\% \\
    \hline
    \end{tabular}
    \label{tab:stats_home_charger}
\end{table}

\begin{table}[H]
    \centering
    \caption{Distribution of dwellings with PV installation \cite{brugelEnergiesRenouvelablesRegion2024,vlaamsenergie-enklimaatagentschapZonnepanelenVlaanderen2024,lallemand16MenagesWallons2023}}
    \small
    \begin{tabular}{|c | c | c | c||c|}
    \hline
     & Brussels & Flanders & Wallonia & Total \\
    \hline
    With PV & 0.2\% & 19.2\% & 6.4\% & 25.8\% \\
    \hline
    Without PV & 10.4\% & 38.2\% & 25.6\% & 74.2\% \\
    \hline
    \end{tabular}
    \label{tab:stats_pv}
\end{table}

\begin{table}[H]
    \centering
    \caption{Distribution of dwellings per construction type \cite{vannesteLogementBelgique2007}}
    \small
    \begin{tabular}{|c | c | c | c||c|}
    \hline
     & Brussels & Flanders & Wallonia & Total \\
    \hline
    Terraced house & 1.3\% & 11.7\% & 8.4\% & 21.4\% \\
    \hline
    Semi-detached house & 0.2\% & 10.6\% & 5.2\% & 16.0\% \\
    \hline
    Detached house & 0.1\% & 18.6\% & 10.1\% & 28.8\% \\
    \hline
    Apartment unit & 9.0\% & 16.5\% & 8.3\% & 33.8\% \\
    \hline
    \end{tabular}
    \label{tab:stats_construction_type}
\end{table}

\begin{table}[H]
    \centering
    \caption{Distribution of dwellings by insulation level \cite{recticelinsulationBarometreLisolation20202020}}
    \small
    \begin{tabular}{|c | c | c | c||c|}
    \hline
     & Brussels & Flanders & Wallonia & Total \\
    \hline
    No insulation & 1.3\% & 7.8\% & 11.5\% & 20.6\% \\
    \hline
    Roof & 5.6\% & 16.5\% & 15.5\% & 37.6\% \\
    \hline
    Wall and roof & - & 10.7\% & 0.2\% & 10.9\% \\
    \hline
    Ground and roof & 0.1\% & - & - & 0.1\% \\
    \hline
    Ground, wall and roof & 3.6\% & 22.4\% & 4.8\% & 30.8\% \\
    \hline
    \end{tabular}
    \label{tab:stats_insulation}
\end{table}

\subsection{Dwelling thermal and heat pump models}
\label{sec:thermal_data_and_models}

The thermal models are created using the Openstudio-HPXML
workflow \cite{horowitzOpenStudioHPXMLWorkflow2024}. This workflow takes as inputs the various characteristics of the dwelling, in this case construction type, insulation level and surface area, and returns and .idf file containing a detailed thermal model of the dwelling with a properly sized heat pump that can then be used for Energyplus \cite{crawleyEnergyPlusCreatingNewgeneration2001} simulations (see \cref{fig:dwelling_flow_chart}).

The roof and ground are insulated with an R-value of 6.4 m²K/W and the walls with a value of 7.3 m²K/W following the values from \citet{recticelinsulationBarometreLisolation20202020}.
The uninsulated R-values are averages of typical Belgian building constructions: 0.5 m²K/W for the walls, 0.3 m²K/W for the roof and 0.4 m²K/W for the ground, as reported by \citet{energie+CoefficientTransmissionThermique2007}.

The windows U-values correspond to single glazing (5.8 W/m²K), double glazing (2.8 W/m²K) and triple glazing (0.6 W/m²K) values from \citet{energuide.beDoubleGlazingTriple2025}.

The air tightness values are taken from \citet{lavergeAirtightnessAssessmentSingle2014} and correspond to the median air tightness values for Belgian houses:
\begin{itemize}
    \item Constructed between 1990 and 1995 (9.3 ACH\textsubscript{50})
    \item Constructed between 2007 and 2010 and randomly sampled (5.5 ACH\textsubscript{50})
    \item Constructed between 2006 and 2013 and sampled from engaged owners that reported their air tightness levels (2.9 ACH\textsubscript{50})
    \item With high energy performance  (1.4 ACH\textsubscript{50})
\end{itemize}

The link between the insulation level, the window U-values and the air tightness is shown in \cref{tab:window_and_tightness}.

\begin{table}[H]
    
    \centering
    \caption{Window U-values and air tightness values}
    \begin{tabular}{|c | c | c |}
    \hline
    Insulation level & Window U-value & Air tightness \\
    \hline
    No insulation & 5.8 W/m²K & 9.3 ACH\textsubscript{50} \\
    \hline
    Roof & 2.8 W/m²K & 5.5 ACH\textsubscript{50} \\
    \hline
    Double insulation & 2.8 W/m²K & 2.9 ACH\textsubscript{50} \\
    \hline
    Ground, wall and roof & 0.6 W/m²K & 1.4 ACH\textsubscript{50} \\
    \hline
    \end{tabular}
    \label{tab:window_and_tightness}
\end{table}

Since only ranges of surface areas are given in the statistics from \cref{tab:stats_surface_area}, each range is assigned a specific value as input to the Openstudio-HPXML workflow.
The surface area is set to 25m\textsuperscript{2} for the dwellings of <35m\textsuperscript{2}, 150m\textsuperscript{2} for the dwellings of >124m\textsuperscript{2} and the average of the upper and lower bounds for the other surface area categories (e.g. 70m\textsuperscript{2} for the 55-84m\textsuperscript{2} range).

The dwelling's heating and cooling are managed by an air-to-air heat pump. This type of heat pump was selected due to its higher sales volume compared to air-to-water and geothermal heat pumps in Belgium \cite{eurobserver22ndReportState2023}.
The heat pump is modelled with a primary heating coil and a primary cooling coil, both operating via a compressor. Additionally, it includes a backup electric resistance heating coil that activates only when the primary heating coil cannot meet the heating demand. The heat pump is also modelled with a resistive defrost heater to prevent ice formation.
The heat pump is controllable by setting a heating and cooling set point temperature.

The historical weather data for Brussels is used in all simulations and was downloaded using the Weather Data Downloader from \citet{oikolabWeatherDataDownloader2025}. The obtained weather data file is in .epw format.

The simulation is conducted in EnergyPlus using the generated .idf thermal model file and the .epw Brussels weather data file as inputs. \Cref{sec:heat_pump_control} details how the heat pump is controlled during the simulation.

\subsection{Domestic hot water}
\label{sec:hot_water}

The domestic hot water system is modelled as a storage tank following the equations from \citet{dolanDevelopmentMonteCarlo1996} with a hot water consumption profile extrapolated from the \citet{athanasiouDAIADOpenWater2014} dataset.

The hot water tank is modelled using the equations presented in \cref{eq:hot_water_params,eq:tank_temp_change}:
\begin{equation}
\begin{aligned}
    &C = \rho^w c^w V  \\
    &G = \frac{SA}{R} \\
    &B_t =  \rho^w c^w F_t \\
    &R'_{t} = \frac{1}{G + B_t} \\
    &D_{t} = \exp\left(-\frac{\Delta t}{R'_{t} C}\right)
\end{aligned}
\label{eq:hot_water_params}
\end{equation}

where:

\begin{tabular}{ll}
    $C$ [J/K]               & tank heat capacity \\
    $\rho^w$ [kg/m$^3$]     & water density (1000 kg/m$^3$) \\
    $c^w$ [J/kg/K]          & specific heat capacity of water (4186 J/kg/K) \\
    $V$ [m$^3$]             & tank volume \\
    $G$, $B_t$, $R'_{t}$, $D_t$ & intermediate parameters and variables \\
    $SA$ [m$^2$]            & tank surface area \\
    $R$ [m$^2$K/W]          & tank insulation $R$-value (used: 10 m$^2$K/W) \\
    $F_t$ [m$^3$/s]         & water flow rate at time $t$ \\
    $\Delta t$ [s]          & time step duration \\
\end{tabular}

\begin{equation}
    T^{tank}_{t+1} = D_{t} \underbrace{T^{tank}_{t}}_{\text{init temp}} + \left( \underbrace{G R'_{t} T^{room}}_{\text{loss to room}}+ \underbrace{B_t R'_{t} T^{inlet}_{t}}_{\text{loss inlet water}} + \underbrace{\eta^{heat} P_t R'_{t}}_{\text{gain heater}} \right) (1 - D_{t})
\label{eq:tank_temp_change}
\end{equation}

where:

\begin{tabular}{ll}
    $T^{tank}_t$ [$^\circ$C] & mean tank temperature at time $t$ \\
    $T^{room}$ [$^\circ$C]   & ambient room temperature (used: 15 $^\circ$C) \\
    $T^{inlet}_t$ [$^\circ$C] & inlet (mains) water temperature at time $t$ \\
    $\eta^{heat}$ [--]             & heater efficiency (used: 1) \\
    $P_t$ [W]               & heater power at time $t$ \\
\end{tabular}

The \citet{athanasiouDAIADOpenWater2014} dataset comprises two and a half years of hourly water consumption data from 1\,007 households in Alicante.
\citet{meirelesDomesticHotWater2022} studied the water consumption of 9\,206 Belgian apartments and gave a month dependant ratio between total water consumption and hot water consumption. The mean of the monthly ratios is 32.3\% with a standard deviation of 2.5\% between the months.
The water consumption profiles from the \citet{athanasiouDAIADOpenWater2014} dataset are converted to hot water consumption profiles by multiplying each month's data by the corresponding monthly hot-to-total water ratio.

The study also gives statistical links between water consumption and the number of bedrooms in the apartments. 
A probability distribution is created linking the number of bedrooms to an estimated average hot water consumption in order to link a dwelling with a hot water consumption profile.

To sample the daily hot water consumption given a number of bedrooms in a dwelling, we use an inverse Gaussian distribution. 
An inverse Gaussian distribution is defined by its mean, shape and location parameters.
Each bedroom count gets its own distribution, with the mean value taken directly from \citet{meirelesDomesticHotWater2022}. The study also provides overall statistics (median, standard deviation, skewness, kurtosis) for hot water use across all apartments.

\begin{figure}
    \centering
    \includegraphics[width=\columnwidth]{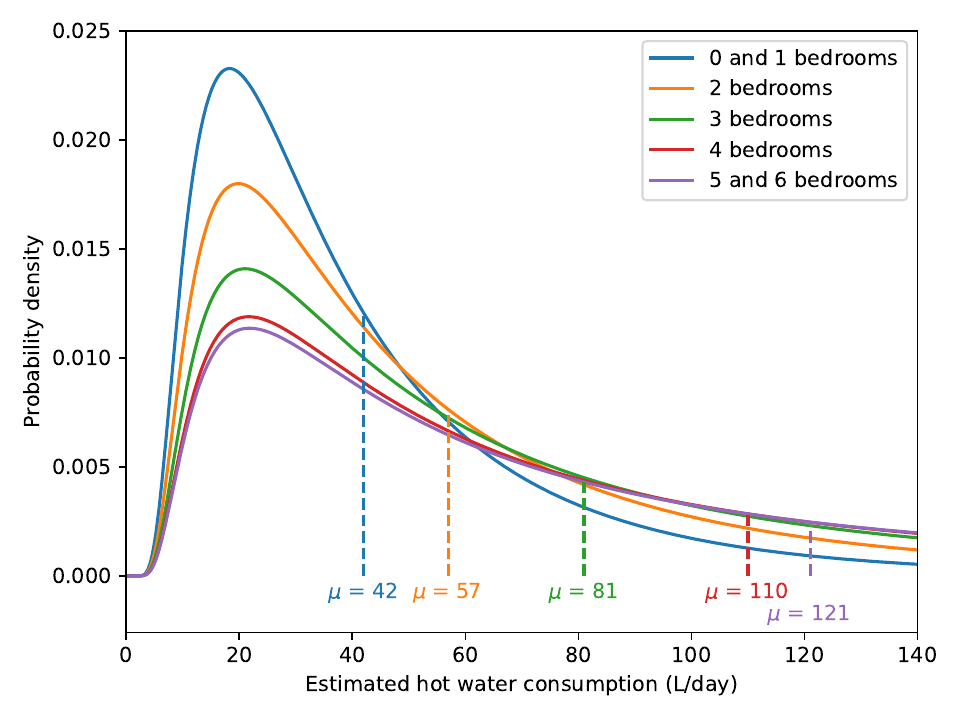}
    \caption{Probability density function of the estimated average daily hot water consumption per number of bedrooms in a dwelling. The mean value of the distribution is indicated by the vertical dashed line.}
    \label{fig:hot_water_distribution}
    \Description{Probability density function of the estimated average daily hot water consumption per number of bedrooms in a dwelling. The mean value of the distribution is indicated by the vertical dashed line.}
\end{figure}

We assume the same shape and location parameters for all distributions. These parameters are fitted so the combination of distributions matches the median value reported in the original dataset. The obtained shape and location parameters are 67.68 and 0 respectively. The resulting probability density functions for each bedroom count are shown in \cref{fig:hot_water_distribution}.

The mean and median of the combination of distributions is therefore the same as the original dataset but the standard deviation (77.2 for dataset vs 65.5 for distribution), skewness (61.3 for dataset vs 5.4 for distribution) and kurtosis (11.3 for dataset vs 66.6 for distribution) are still different. Knowing the exact distribution for each bedroom count would require the original data which is not available.

The hot water consumption profile for a dwelling is determined by sampling from the inverse Gaussian distribution that matches the number of bedrooms in the dwelling. The profile selected is the one with the average daily hot water consumption closest to the sampled value.

In addition to the hot water consumption profile, the model from \citet{dolanDevelopmentMonteCarlo1996} need a temperature for the incoming cold water from the mains.
\citet{fuentesReviewDomesticHot2018} established that a sinusoidal function, which provides the mains water temperature as a function of the month is a good approximation. 
The upper and lower bounds of the sinusoidal function are set to 11\,°C and 16\,°C respectively \cite{v.sebastienMachinesLaverConsommentelles2020}.

The water heater is modelled as an electric resistance heater with a heater efficiency ($\eta^{heater}$) of 1. The volume of the storage tank is three times the expected average daily hot water consumption determined by the number of bedrooms to account for peak consumption periods (so 146 litres for a dwelling with 1 bedroom and 363 litres for a dwelling with 6 bedrooms).

\subsection{Electric vehicle charger}
\label{sec:ev_charger}

The dwellings sampled to have a garage and a car (see \cref{tab:stats_home_charger}) are equipped with an EV charger.
Each simulation uses a real user charging profile, which is a record of every home charging session information that occurred over the year. These profiles are sourced from \citet{sorensenElectricVehicleCharging2024}.
This dataset includes the arrival time, departure time and charged energy of each charging session of EVs from 267 users charging at 12 residential locations in Norway.
The dataset also provides assumed arrival and departure state of charge, battery sizes and maximum charging powers for each charging session \cite{sorensenMethodGeneratingComplete2023}.
All charging profiles containing more than 600 days of data were included in this study, resulting in a total of 16 profiles.
Each dwelling with an EV charger is randomly assigned one of these 16 charging profiles.

\begin{table}
    \centering
    \caption{Summary statistics of the 16 EV charging profiles used in this study.}
    \small
    \begin{tabular}{|l|c|c|}
        \hline
        Statistic & Mean & Std. dev. \\
        \hline
        Total energy charged per profile per year & 2\,414 kWh & 1\,464 kWh \\
        Number of sessions in a year per profile & 193.6 & 131.2 \\
        Session duration across all profiles & 12h16min & 11h12min \\
        Session SOC charged across all profiles  & 31.7\% & 20.0\% \\
        Std. dev. of session duration per profile & 9h46min & 5h32min \\
        Std. dev. of session SOC charged per profile & 17.3\% & 2.6\% \\
        \hline
    \end{tabular}
    \label{tab:ev_charging_profile_stats}
\end{table}

\Cref{tab:ev_charging_profile_stats} summarizes the variability in the EV charging profiles. 

On average, each EV charges by 2\,414 kWh per year at home. Assuming an EV efficiency of 0.191 kWh/km \cite{evdatabaseEnergyConsumptionFull2025}, this equates to an annual driving distance of 12\,638 km. This figure is close to the average yearly travelled distance by Belgian cars of 14\,420 km \cite{carpassRapportAnnuel20242025}, especially considering that not all users charge their EV exclusively at home.

On average, each profile has about 194 charging sessions per year, but this number varies widely between users (standard deviation: 131). The typical charging session lasts 12 hours and 16 minutes, with a standard deviation of 11 hours and 12 minutes, showing that some sessions are much shorter or longer than others. The average state of charge (SOC) charged per session is 31.7\%, but again, this varies a lot (standard deviation: 20\%). Even when looking at individual profiles, the average standard deviation within a user's sessions remains high at 9 hours and 46 minutes for session duration, and at 17.3\% for charged SOC. 

In summary, the average charged energy per year across the users is typical for Belgian drivers, and EV charging behaviour is highly variable both between users and from session to session for the same user. This highlights potential challenges in predicting this behaviour for energy management systems.

The model of the charged EV is defined by \cref{eq:ev_charging_reliability}:

\begin{equation}
    \label{eq:ev_charging_reliability}
    \text{SOC}_{t+1} = \text{SOC}_{t} + \frac{\eta^{charge} \times \text{P}^{\text{charge}}_{t} \times \Delta t}{\text{E}^{\text{batt}}}
\end{equation}

where:

\begin{tabular}{ll}
    $\text{SOC}_{t}$ [--]           & State of charge of the EV battery at time $t$ \\
    $\text{P}^{\text{charge}}_{t}$ [kW] & Power of the charger at time $t$ \\
    $\Delta t$ [h]                  & Time step duration \\
    $\text{E}^{\text{batt}}$ [kWh]    & Energy capacity of the battery \\
    $\eta^{\text{charge}}$ [--]     & Charging efficiency \\
\end{tabular}

The maximum charging power provided in the dataset for the 16 charging profiles is 3.7 kW. We are aware that home chargers in Belgium generally have a higher maximum power (7.4 kW or 11 kW) but we chose to keep the dataset's maximum charging power because the maximum power of the charger could influence the user's charging profile unpredictably (e.g. shorter connection times).

The charging efficiency $\eta^{\text{charge}}$ is set to 88\% and the departure SOC is always 95\% which are the assumptions made by \citet{sorensenMethodGeneratingComplete2023} for the dataset.

\subsection{Photovoltaic panels}
\label{sec:pv_panels}

The PV panels are simulated using real Belgian PV production data available on a 5-minute granularity from the PVoutput.org website \cite{pvoutput.orgPVOutputorg2025}.
A total of 380 PV profiles were downloaded from the website, representing all installations with a capacity below 10 kW that recorded measurements for at least 95\% of the days in both 2022 and 2023.
Each dwelling with PV panels is randomly assigned a PV profiles from its corresponding region out of the 380 profiles.

\subsection{Electricity consumption}
\label{sec:consumption_profiles}

The electricity consumptions are simulated using real house electricity consumption profiles available on a 2-minute granularity for houses located in Hamelin, Germany \cite{schlemmingerDatasetElectricalSinglefamily2022}.
The heating and cooling electricity consumption is measured separately from the rest of the electricity consumption, this study only uses the rest of the electricity consumption. 

Out of the 38 profiles available in the dataset, this study used the 22 profiles with data available for at least 98\% of the time steps for the years 2019 and 2020. 
Since all missing data must be imputed for the simulation, this threshold ensures that only a minimal amount of data requires imputation.

The 22 consumption profiles have an average annual electricity consumption of 2\,833 kWh, with a standard deviation of 862 kWh. For comparison, the average Flemish household consumed 2\,534 kWh according to Vlaamse Nutsregulator, while the European average reported by Eurostat is 3\,500 kWh \cite{vlaamsenutsregulatorEnergieverbruik2024}. These profiles are therefore reasonably representative of typical household yearly consumption, with the added benefit of being based on real, high-resolution measurements.

\subsection{Day-ahead electricity prices}

\Cref{tab:day_ahead_price_stats} summarizes the  electricity prices for the Belgian day-ahead market ($Vd_t$) from 2020 to 2025.
The table highlights significant year-to-year variation. Notably, 2022 stands out as an exceptional year with unusually high prices, whereas 2023 falls within a more typical range observed over recent years.

\begin{table}
    \centering
    
    \caption{Electricity prices for the Belgian day-ahead market for years 2020 to 2025. All values are in €/kWh \cite{entso-eENTSOETransparencyPlatform2025}.} 
    \begin{tabular}{|c|c|c|c|c|}
        \hline
        Year & Mean & Std. dev. & Min & Max \\
        \hline
        2020 & 0.032 & 0.017 & -0.115 & 0.200 \\
        2021 & 0.104 & 0.079 & -0.070 & 0.620 \\
        2022 & 0.245 & 0.135 & -0.100 & 0.871 \\
        2023 & 0.097 & 0.046 & -0.120 & 0.330 \\
        2024 & 0.070 & 0.0430 & -0.140 & 0.566 \\
        2025 & 0.085 & 0.0482 & -0.462 & 0.518 \\
        \hline
    \end{tabular}
    \label{tab:day_ahead_price_stats}
\end{table}

\section{MPC model}
\label{sec:mpc_model}

\subsection{MPC mathematical formulation}

The MPC model is formulated as a mixed-integer linear programming problem.

The dwelling's thermal model is approximated using the resistor-capacitor analogy for thermal modelling \cite{madsenEstimationContinuoustimeModels1995,drgonaAllYouNeed2020}. \Cref{eq:mpc_thermal_model} shows the implementation of this model in the MPC. EnergyPlus does not allow to control the power of the heat pump directly so the control setpoint used is the temperature of the model at the next time step ($T^{in}_{t+1}$).

\begin{equation}
C^{dwell}\frac{T^{in}_{t+1}-T^{in}_{t}}{\Delta t} = \frac{T^{out}_{t}-T^{in}_{t}}{TR}+Q^{pump}_{t} + gAQ^{sun}_{t}
\label{eq:mpc_thermal_model}
\end{equation}

where:

\begin{tabular}{ll}
    $C^{dwell}$ [J/K] & Thermal capacitance of the dwelling \\
    $T^{in}_{t}$ [K] & Indoor temperature at time step $t$ \\
    $T^{out}_{t}$ [K] & Outdoor temperature at time step $t$ \\   
    $TR$ [K/W] & Thermal resistance of the dwelling \\
    $Q^{pump}_{t}$ [W] & Thermal power of the heat pump at time step $t$ \\
    $gA$ [m²] & Solar aperture \\
    $Q^{sun}_{t}$ [W/m] & Solar irradiance at time step $t$ \\
\end{tabular}

The primary heating and cooling coils and the backup heating coil are modelled linearly using a fixed efficiency conversion between the electric power the thermal power as shown in \cref{eq:mpc_heat_pump}. 

\begin{equation}
Q^{pump}_{t} = \eta^{heat}P^{heat}_{t}-\eta^{cool}P^{cool}_{t}+\eta^{backup}P^{backup}_{t}
\label{eq:mpc_heat_pump}
\end{equation}

where:

\begin{tabular}{ll}
    $\eta^{coil}$ [--] & Efficiency of the heat, cool or backup coil 
     \\
    $P^{coil}_{t}$ [W] & \begin{tabular}{@{}l@{}}Electric power of the heat, cool or backup coil at \\time step $t$\end{tabular} \\
\end{tabular}

The total electric power of the heat pump $P_{t}^{pump}$ is modelled as the sum of the electric power of the primary heating and cooling coils and the backup heating coil as shown in \cref{eq:mpc_heat_pump_total}.

\begin{equation}
P^{pump}_{t} = P^{heat}_{t}+P^{cool}_{t}+P^{backup}_{t}
\label{eq:mpc_heat_pump_total}
\end{equation}

The parameters for the thermal capacitance ($C^{dwell}$), thermal resistance ($TR$), solar aperture ($gA$), and coil efficiencies ($\eta^{heat}$, $\eta^{cool}$, $\eta^{backup}$) are calibrated by fitting the MPC model to the detailed simulation results for each dwelling over the year 2022. 

When using the accurate simulation values for $T_t^{out}$, $T_t^{in}$, $Q_t^{pump}$ and $Q_t^{sun}$, the calibrated model from \cref{eq:mpc_thermal_model} predicts the indoor temperature for the next time step $T_{t+1}^{in}$ with a mean absolute error of 0.12\,°C (standard deviation 0.14\,°C) across all dwellings. This outperforms a naive baseline that assumes the indoor temperature remains constant, which yields a mean absolute error of 0.16\,°C (standard deviation 0.31\,°C).
These results indicate that the simulation's thermal dynamics are more complex than those captured by the simplified RC model used in the MPC.

The maximum electric power of each heat pump coil is constrained by the maximum capacities defined by the Openstudio-HPXML workflow for the respective dwelling.
The MPC model does not include the ventilation consumption and the defrosting coil consumption present in the simulation model.

The indoor temperature $T^{in}_{t}$ is constrained to be within the comfort range defined in \cref{sec:heat_pump_control} for the predicted horizon.

The water heater's MPC model closely mirrors the simulation model by implementing \cref{eq:tank_temp_change} and enforcing heating at maximum power for 15 minutes whenever the tank temperature falls below 55\,°C.
The main difference is that, in the MPC model, the tank temperature is strictly limited to a maximum of 65\,°C, rather than allowing it to exceed this value and simply disabling heating in the following time step.

The EV charger is modelled using \cref{eq:ev_charging_reliability} for the EV charging and the charging session is constrained to charge a given amount of energy in a given time period.
This exactly matches the simulation model of the charger.

The total electric power of the dwelling is the sum of the different assets as shown in \cref{eq:mpc_total_power}.

\begin{equation}
    P^{total}_{t} = P^{pump}_{t}+P^{water}_{t}+P^{EV}_{t}-P^{PV}_{t}+P^{load}_{t}
\label{eq:mpc_total_power}
\end{equation}

Overall, the MPC model provides a close approximation of the simulation model.

\subsection{Objective function}

To calculate the objective function, the energy offtake $Eo_t$ and injection $Ei_t$ are calculated using the total electric power $P^{total}_{t}$ following \cref{eq:mpc_energy_offtake,eq:mpc_energy_injection}.

\begin{equation}
    Eo_t = 
    \begin{cases}
        P^{total}_{t} \Delta t, & \text{if } P^{total}_{t} \geq 0 \\
        0, & \text{otherwise}
    \end{cases}
    \label{eq:mpc_energy_offtake}
\end{equation}
\begin{equation}
    Ei_t = 
    \begin{cases}
        -P^{total}_{t} \Delta t, & \text{if } P^{total}_{t} < 0 \\
        0, & \text{otherwise}
    \end{cases}
    \label{eq:mpc_energy_injection}
\end{equation}

Model predictive control uses an adapted version from \cref{eq:tot_cost_usefull} as its objective function without the yearly cost, which cannot be optimized, and where the peak cost increase is directly reflected on the prediction horizon (see \cref{eq:mpc_obj_relevance}).

\begin{equation}
O^{MPC}=\underbrace{\sum_{t\in H}{(Eo_tVo_t-Ei_tVi_t)}}_{\text{day-ahead}}+\underbrace{\sum_{t\in H}Eo_tVf}_{\text{offtake extras}}+\underbrace{Vp \times \max_{{t\in H}} (\frac{Eo_t}{\Delta t}, Pp)}_{\text{peak}}
\label{eq:mpc_obj_relevance}
\end{equation}

where:

\begin{tabular}{ll}
    $H$ & Set of time steps in the prediction horizon \\
    $Pp$ [kW] & \begin{tabular}{@{}l@{}}Previous grid peak power for that month \\ (minimum 2.5 kW)\end{tabular} \\
\end{tabular}

\section{TreeC training}

\subsection{Training process and pruning}

The training of TreeC is performed by evaluating the performance of different decision tree based EMSs on the simulation of the year 2022. The enforced charging mechanism described in \cref{sec:ev_charger_control} is not used during the TreeC training simulations to penalize EMSs that do not adequately charge the EV.

Two changes have been made to the original TreeC method described in \citet{ruddickTreeCMethodGenerate2025}:

\begin{itemize}
\item The covariance matrix adaptation evolution strategy was replaced by the metaheuristic algorithm particle swarm optimization \cite{kennedyParticleSwarmOptimization1995} as it performed better in \citet{ruddickRealworldValidationSafe2024}.
This study used the canonical particle swarm optimization algorithm \cite{poliParticleSwarmOptimization2007} implementation from pygmo \cite{biscaniParallelGlobalMultiobjective2020}.
\item After training, the best decision tree is pruned by iteratively removing leaves whose removal does not degrade the objective value by more than 5\,€ compared to the unpruned tree. This threshold is selected to balance performance and interpretability: it is low enough to avoid sacrificing meaningful cost savings, yet high enough to eliminate leaves that contribute minimally to overall performance.
The least used leaves are tested for removal first. 
\end{itemize}

The TreeC EMS is trained in simulation with the particle swarm optimization algorithm over 300 generations and a population of 200 individuals.
This means 60\,000 decision tree based EMSs are evaluated in simulation on the 2022 training data and the best one is then selected.
The population size of 200 was chosen as it obtained better results for the real-world and non-unimodal benchmarks of the IEEE Competitions in Evolutionary Computation compared to the more historically implemented population sizes of 20-50 individuals \cite{piotrowskiPopulationSizeParticle2020}.
To reduce the risk of converging to a local optimum, five independent EMSs are trained and pruned for each dwelling. The EMS with the best performance among the five is then selected to be evaluated on the 2023 test data.

The TreeC training was performed on a high performance computer
composed of Intel\textregistered{} Xeon\textregistered{} Gold, AMD EPYC\texttrademark{} 9384X and AMD EPYC\texttrademark{}  9535 processors. Each training used 100 cores and evaluated 300\ 000 EMSs (60\,000 per metaheuristic optimization $\times$ 5 to avoid local optimums). On average a training took 5 days and 20 hours to complete with each EMS evaluation taking on average 34 seconds on a single core corresponding to the simulation of the dwelling for one year.
\subsection{Objective function}
\label{sec:treec_objective_function}

The total electricity cost $Ctot$ is calculated using \cref{eq:tot_cost_usefull}.

Discomfort is measured in Kh (Kelvin-hours), which represents how long and by how much the indoor temperature deviates from the comfort range. For example, being 2 °C above the comfort range for 1 hour results in 2 Kh of discomfort.

The discomfort above the upper bound of the comfort range $\delta^{up}_t$ and below the lower bound $\delta^{low}_t$ are calculated using \cref{eq:discomfort_cost_up,eq:discomfort_cost_low}.

\begin{equation}
     \delta^{up}_t = \int_{0}^{\Delta t} \max\left( \frac{T_{t+1} - T_{t}}{\Delta t} x + (T_t - T^{up}_t),\, 0K \right) dx
     \label{eq:discomfort_cost_up}
\end{equation}

\begin{equation}    
    \delta^{low}_t = \int_{0}^{\Delta t} \max\left( \frac{T_{t} - T_{t+1}}{\Delta t} x + (T^{low}_t - T_t),\, 0K \right) dx
    \label{eq:discomfort_cost_low}
\end{equation}

where:

\begin{tabular}{ll}
    $T_{t}$ [K] & Indoor temperature at time step $t$ \\
    $\Delta t$ [h] & Length of the time step \\
    $T^{up}_t$ [K] & Upper bound of the comfort range at time step $t$ \\
    $T^{low}_t$ [K] & Lower bound of the comfort range at time step $t$ \\
\end{tabular}

The total discomfort cost $C\delta$ is calculated by \cref{eq:discomfort_cost_usefull}.

\begin{equation}
    C\delta = 
   max(0\text{€},
        w1\times(\sum_{t \in T} (\delta^{up}_t + \delta^{low}_t)-30 Kh))
    \label{eq:discomfort_cost_usefull}
\end{equation}

where:

\begin{tabular}{ll}
    $C\delta$ [€] & Discomfort cost \\
    $w_1$ [€/Kh] & Penalty weight for discomfort cost (5 €/Kh) \\
\end{tabular}

A discomfort penalty is only applied if the total annual discomfort exceeds 30 Kh, which corresponds to a low level of discomfort over the year. For any discomfort above this threshold, a penalty of 5\,€/Kh is imposed, strongly incentivizing the EMS to maintain thermal comfort.

The cost for missed EV charging energy $Ccharg$ is calculated using \cref{eq:charg_cost_usefull}.

\begin{equation}
    Ccharg = \sum_{s \in S} 
    \begin{cases}
        w_2 \times E^{miss}_s, & \text{if } E^{miss}_s \leq E^{5\%} \\
        w_2 \times E^{5\%} + w_3 \times (E^{miss}_s - E^{5\%}), & \text{otherwise}
    \end{cases}
    \label{eq:charg_cost_usefull}
\end{equation}

where:

\begin{tabular}{ll}
    $Ccharg$ [€] & Penalty cost for missed EV charging energy \\
    $S$ & Set of all charging sessions \\
    $E^{miss}_s$ [kWh] & Missed energy in session $s$ \\
    $E^{5\%}$ [kWh] &\begin{tabular}{@{}l@{}}
         Energy corresponding to charging the EV by \\5\% SOC
    \end{tabular}   \\
    $w_2$ [€/kWh] & \begin{tabular}{@{}l@{}}Penalty weight for missed energy up to $E^{5\%}$ \\ (0.5 €/kWh)\end{tabular} \\
    $w_3$ [€/kWh] & \begin{tabular}{@{}l@{}}Penalty weight for missed energy exceeding \\ $E^{5\%}$ (2 €/kWh)\end{tabular} \\
\end{tabular}

A missed energy event occurs when the simulated EMS charges less energy than what was originally delivered in the dataset for a given charging session. In other words, it is the shortfall between the original and the actually delivered energy by the EMS. 

Missing up to 5\% of the EV's SOC in a charging session is undesirable but not critical, so a moderate penalty of 0.5\,€/kWh is applied reflecting an exceptionally high price of electricity (see \cref{tab:day_ahead_price_stats}). However, missing more than 5\% SOC is considered a significant problem, and a much higher penalty of 2\,€/kWh is imposed to strongly discourage such outcomes.

The total objective function $O$ of the TreeC EMS is calculated by summing the different costs from \cref{eq:tot_cost_usefull,eq:discomfort_cost_usefull,eq:charg_cost_usefull} as shown in \cref{eq:treec_obj_usefull}.

\begin{equation}
    O^{TreeC} = Ctot +  C\delta +  Ccharg
    \label{eq:treec_obj_usefull}
\end{equation}

\end{document}